\documentclass[reprint]{revtex4-2}
\usepackage[dvipsnames]{xcolor}
\usepackage{graphicx}
\usepackage{amsmath}
\usepackage{tablefootnote}
\usepackage{amssymb}
\usepackage{slashed}
\usepackage{hyperref}
\usepackage{enumerate}
\usepackage{tabularx}
\usepackage{bbold}

\usepackage{csvsimple}

\usepackage{lipsum}
\usepackage[normalem]{ulem}

\graphicspath{ {./Images/} }

\newcommand{\Mmax}{{M_\text{max}}}

\begin{document}

\title{A Bayesian Inference of Hybrid Stars with Large Quark Cores}

\author{Milena Albino}
\email{milena.albino@student.uc.pt}
\affiliation{Department of Physics, CFisUC, University of Coimbra, P-3004 - 516 Coimbra, Portugal}

\author{Tuhin Malik}
\email{tm@uc.pt}
\affiliation{Department of Physics, CFisUC, University of Coimbra, P-3004 - 516 Coimbra, Portugal}

\author{Márcio Ferreira}
\email{marcio.ferreira@uc.pt}
\affiliation{Department of Physics, CFisUC, University of Coimbra, P-3004 - 516 Coimbra, Portugal}

\author{Constança Providência}
\email{cp@uc.pt}
\affiliation{Department of Physics, CFisUC, University of Coimbra, P-3004 - 516 Coimbra, Portugal}

\date{\today}

\begin{abstract}
Neutron stars (NSs) are interesting objects capable of reaching densities
unattainable on Earth.
The properties of matter under these conditions remain a mystery.
Exotic matter, including quark matter, may be present in the NS core.
In this work, we explore the possible compositions of NS cores, {in particular, the possible existence of large quark cores}.
We use the Relativistic Mean Field (RMF) model with nonlinear terms for the hadron phase
and the Nambu–Jona-Lasinio (NJL) model
and Mean Field Theory of Quantum Chromodynamics (MFTQCD) for the quark phase.
Through Bayesian inference, we obtain different sets of equations: four sets with hybrid equations
and one set with only the hadron phase.
We impose constraints regarding the  properties of nuclear matter, X-ray observational data from NICER, gravitational wave data from the binary neutron star merger GW170817,
perturbative QCD (pQCD) calculations, and causality.
The MFTQCD allows for a phase transition  to quark matter at low densities, just above saturation density, while for the NJL sets, the phase transition occurs above twice the  saturation density. 
 As a result, the MFTQCD model predicts the presence of quark matter in the inner core of 1.4$\text{M}_\odot$ NSs, while NJL models suggest a low probability of quark matter in the interior of a 1.4 $\text{M}_\odot$ NS. Both models predict the existence of quark matter in 2$\text{M}_\odot$ NSs. The slope of the mass-radius curve has been shown to carry information about the presence of quark matter. In particular, a positive slope at 1.8  $\text{M}_\odot$ indicates the presence of non-nucleonic matter.
A hybrid star with a stiff quark equation of state could explain a larger radius in more massive stars, such as two solar mass stars, compared to canonical NSs.
\end{abstract}

\maketitle

\section{Introduction}
  The internal composition of neutron stars (NSs)
  remains one of the most significant open questions in nuclear astrophysics.
  These compact objects, with masses of approximately {1.2--2.0 solar masses ($\text{M}_{\odot}$)}
  concentrated within a radius of only 10-14 km,
  represent the densest observable matter in the universe \cite{Glendenning_2000,Rezzolla:2018jee}.
  At such extreme densities, exceeding several times the nuclear saturation density
  ($\rho_0 \approx 2.7 \times 10^{14}$ g cm$^{-3}$)
  our understanding of matter's behavior becomes increasingly uncertain
  due to the limitations of terrestrial experiments and first-principles calculations.
  NSs are believed to provide a natural laboratory
  for studying exotic high-density phases of QCD,
  such as the neutron superfluid phase \cite{Rezzolla:2018jee,Lovato:2022vgq}.
  The outer cores of NSs are so dense and
  thick that electromagnetic signals cannot escape,
  while theoretical calculations of the inner cores
  are hindered by the limitations of first principles lattice  Quantum Chromodynamics (QCD).

  A particularly intriguing possibility is that
  NSs may undergo a phase transition
  from hadronic matter to quark matter in their inner cores,
  forming what are known as hybrid stars \cite{Glendenning_2000,Schertler:1999xn,Menezes:2003xa,Alford:2006vz,Albino:2024ymc,Blacker:2024tet}.
  Quantum Chromodynamics (QCD), the fundamental theory of strong interactions,
  predicts a transition from confined hadronic matter
  to a deconfined quark-gluon plasma at sufficiently high densities or temperatures.
  While numerical simulations of QCD at vanishing baryonic chemical potential
  indicate a smooth crossover transition at
  a temperature of $T \approx 154.9$ MeV \cite{HotQCD:2018pds},
  the nature of this transition at
  the high densities and relatively low temperatures relevant for NS interiors
  remains an open question.
  Some studies suggest that finite surface tension effects
  can lead to mixed phase states with different geometric shapes (known as ``pasta'' phases),
  potentially inducing a smooth phase transition \cite{Glendenning:2001pe,Maruyama:2007ey,Pradhan2023}.
  Depending on the nature of the phase transition,
  a third family of stable, compact stars (twin stars)
  with different radii compared to normal NSs may appear,
  providing a unique observational signature of the hadron-quark transition \cite{Benic:2014jia,Pradhan2023}.

  Recent breakthroughs in multi-messenger astronomy
  have opened unprecedented opportunities to probe the properties of supranuclear matter.
  The detection of gravitational waves (GWs) from binary NS mergers,
  beginning with GW170817 \cite{LIGOScientific:2017vwq,LIGOScientific:2018cki,Chatziioannou2019},
  coupled with electromagnetic observations \cite{LIGOScientific:2017zic},
  has provided valuable constraints on the NS equation of state (EOS).
  Additionally, precise mass and radius measurements from
  NASA's Neutron Star Interior Composition Explorer (NICER) mission
  have further constrained the possible EOS models \cite{Malik:2022zol,Huang:2023grj,Pradhan2023}.
  The accuracy of pulsar timing,
  comparable to that of atomic clocks (one part in 10$^{15}$),
  allows for the indirect detection of GWs from binary NS merger events
  and provides a window for exploring phase transitions
  occurring inside a pulsar core \cite{Bagchi2023}.
  Notably, the GW signal emitted
  during the final orbits of colliding NSs contains imprints of
  the tidal deformability parameter $\Lambda$,
  which can be related to the properties of dense matter
  in terms of the EOS \cite{Bauswein:2017vtn,Most:2018hfd,Raithel2022}.

  These observational advances have motivated a renewed
  theoretical effort to develop more sophisticated EOS models
  that can account for potential phase transitions
  while remaining consistent with observational constraints.
  In this context, Bayesian inference has emerged as a powerful framework
  for parameter estimation and model selection in astrophysics \cite{Wysocki2020}.
  It provides a natural way to incorporate prior knowledge,
  handle uncertainties, and update our beliefs based on new observations.

  In this work, we present a comprehensive Bayesian analysis
  of the EOS for hybrid stars,
  simultaneously sampling the parameters of both the hadron and quark matter phases.
  For the hadron phase, we employ the Relativistic Mean Field (RMF) theory with non-linear terms,
  which has been widely used to describe nuclear matter and finite nuclei \cite{Sugahara:1993wz,Mueller:1996pm,Lalazissis:1996rd,Horowitz:2000xj,Todd-Rutel:2005yzo,Chen:2014sca,Malik:2023mnx}.
  This approach uses the RMF model to describe the hadron phase of NS matter,
  involving baryons interacting through the exchange of mesons.
  For the quark matter phase, we adopt two models:
  the Nambu--Jona-Lasinio (NJL) model \cite{NJL1,NJL2,Klevansky:1992qe,Hatsuda:1994pi} and
  the Mean Field Theory of QCD (MFTQCD) model \cite{Fogaca:2010mf}.
  The first one incorporates key features of QCD such as
  global symmetries from QCD
  and dynamical chiral symmetry breaking and its manifestations \cite{Hatsuda:1994pi,VOGL1991195,Buballa:1998pr,buballa}.
  The NJL model offers an attractive framework as
  it describes the quark phase with a three-flavor version
  and parameters determined by fitting to various meson and baryon masses \cite{buballa,CamaraPereira:2016chj}.
  The second model is obtained by decomposing the gluon field
  into low- and high-momentum components and
  applying suitable approximations to these fields.
  The MFTQCD EOS exhibits behavior similar to that of the vectorial MIT bag model
  \cite{Lopes:2020btp}
  and can result in an mass-radius diagram consistent with recent observations
  \cite{Franzon:2012in,Albino:2021zml}.
  This combined approach allows us to explore a more complete picture
  of the dense matter in NS cores.

  The existence of phase transitions in NS cores can manifest
  in observable signatures across multiple messengers.
  In the mass-radius diagram,
  a strong first-order phase transition can produce
  disconnected branches of stable configurations,
  leading to the possibility of ``twin stars''
  - NSs with the same mass but different radii \cite{Benic:2014jia,Pradhan2023}.
  During binary NS mergers,
  the hadron-quark phase transition can significantly affect
  the dynamics of the system and the emitted GWs \cite{Bauswein2018,Most:2018eaw,Hammond:2025kki}.

  The transition from hadron to quark matter in our model employs
  the Maxwell construction for phase equilibrium,
  where the transition occurs at a specific pressure
  with a discontinuity in energy density \cite{Glendenning_2000,Blacker2020}.
  While the Gibbs construction allows for a mixed phase where hadronic and quark matter coexist, the Maxwell construction assumes a sharp interface between the two phases
  with equal pressures but different densities.
  A further option would be to construct a phase transition that smoothly connects the equations of the phases
  \cite{Kojo_2015,Hammond:2025kki,Albino:2025oiw}.
  Our choice of the Maxwell construction is motivated by several considerations.
  First, it provides a more conservative estimate of the transition effects,
  as the energy density discontinuity leads to
  more pronounced observational signatures in GW emission \cite{Guo2023}.
  Second, the surface tension at the hadron-quark interface,
  though poorly constrained, is believed to be sufficiently large
  to disfavor the formation of mixed-phase structures in many scenarios \cite{Maruyama:2007ey,Prakash2021}.
  Third, the simplified thermodynamic treatment of the Maxwell construction
  is computationally advantageous for our Bayesian parameter estimation,
  allowing for more extensive sampling of the parameter space.

  By performing this comprehensive Bayesian inference of the hybrid star EOS,
  we aim to address several key questions:
  (1)
  Is it possible to build hybrid EOSs that satisfy current observations and theoretical predictions while containing a large quark core?
  (2) What are the most likely properties of this transition, such as its onset density and strength?
  (3)
  Which properties distinguish purely hadronic from hybrid EOS models?
\\
  
  The article is organized as follows.
  In Sec. \ref{sec:EOS},
  we describe the theoretical framework for the RMF, NJL and MFTQCD models
  and the construction of the hybrid EOS.
  Sec. \ref{sec:Bayesian} outlines our Bayesian methodology,
  including the prior distributions, likelihood function, and computational techniques.
  In Sec. \ref{sec:Results}, we present our results
  on the posterior distributions of model parameters
  and the resulting constraints on the hybrid EOS.
  Sec. \ref{sec:Implications} discusses the implications of our findings
  for NS observations and fundamental nuclear physics.
  Finally, Sec. \ref{sec:Conclusion} summarizes our conclusions
  and outlines directions for future work.

\section{Equations of State} \label{sec:EOS}
  This section describes the models used for the hadron and quark phases.
  Hybrid EOSs are built through Maxwell construction.
  Four sets of hybrid EOSs were generated:
  three using the NJL model for the quark phase with different priors and constraints
  and one using the MFTQCD for the quark phase.
  A fifth set consisting only of nucleonic matter was also generated.
  The hadron phase is described by the RMF model in all cases.

\subsection{Hadron phase}
  The Relativistic Mean Field (RMF) model is used for the hadron phase.
  In this model, nucleon interactions are mediated by the exchange of mesons:
  the scalar-isoscalar meson $\sigma$,
  the vector-isoscalar meson $\omega$, and
  the vector-isovector meson $\varrho$.
  This work also adds non-linear terms.
  The Lagrangian is given by \cite{Malik:2023mnx}:
  \begin{equation}
    \mathcal{L} = \mathcal{L}_N +\mathcal{L}_M +\mathcal{L}_{NL},
  \end{equation}
  where
  \begin{align}
    \mathcal{L}_N &= \bar{\Psi}
      \left[
        \gamma^\mu \left(
          i \partial_\mu -g_\omega \omega_\mu
          -g_\rho \boldsymbol{t} \cdot \boldsymbol{\rho}_\mu
        \right)
      \right. \nonumber \\
      &\quad \left.
        -(m -g_\sigma \phi)
      \right] \Psi,  \\
    \mathcal{L}_M &= \frac{1}{2}
      \left[
          \partial_\mu \sigma \partial^\mu \sigma
          -m_\sigma^2 \sigma^2
      \right] \nonumber \\
      &\quad
        -\frac{1}{4} \omega_{\mu\nu} \omega^{\mu\nu}
        +\frac{1}{2} m_\omega^2 \omega_\mu \omega^\mu \nonumber \\
      &\quad
        -\frac{1}{4} \boldsymbol{\varrho}_{\mu\nu} \cdot \boldsymbol{\varrho}^{\mu\nu}
        +\frac{1}{2} m_\varrho^2 \boldsymbol{\varrho}_\mu \cdot \boldsymbol{\varrho}^\mu, \\
    \mathcal{L}_{NL} &=
      -\frac{1}{3} b m g_\sigma^3 (\sigma)^3
      -\frac{1}{4} c g_\sigma^4 (\sigma)^4
      +\frac{\xi}{4!} g_\omega^4 (\omega_\mu \omega^\mu)^2 \nonumber \\
      &\quad
      +\Lambda_\omega g_\varrho^2 \boldsymbol{\varrho}_\mu \cdot \boldsymbol{\varrho}^\mu
      g_\omega^2 \omega_\mu \omega^\mu,
  \end{align}
  where $\Psi$ represents the Dirac spinor nucleon doublet (proton and neutron)
  with a bare mass $m$, $g_i$ and $m_i$ are the coupling constants and the masses
  of the mesons $i = \sigma, \omega, \varrho$ and
  $\omega_{\mu\nu} = \partial_\mu \omega_\nu -\partial_\nu \omega_\mu$
  and similar  for $\boldsymbol{\varrho}_{\mu\nu}$. The equations of motion for $\sigma$, $\omega$ and $\varrho$ mesons are determined from the Euler-Lagrangian equations:
  \begin{align}
    \sigma = &\frac{g_\sigma}{m_{\sigma,\text{eff}}^2} \sum_i \rho_i^S,
    \\
    &m_{\sigma,\text{eff}}^2 = -m_\sigma^2 -b m g_\sigma^3 \sigma -c g_\sigma^4 \sigma^2, \\
    \omega = &\frac{g_\omega}{m_{\omega,\text{eff}}^2} \sum_i \rho_i,
    \\
    &m_{\omega,\text{eff}}^2 = m_\omega^2
      +\frac{\xi}{3!} g_\omega^4 \omega^2
      +2 \Lambda_\omega g_\varrho^2 g_\omega^2 \varrho^2, \\
    \varrho = &\frac{g_\varrho}{m_{\varrho,\text{eff}}^2} \sum_i t_3 \rho_i,
    \\
    &m_{\varrho,\text{eff}}^2 = m_\varrho^2
      +2 \Lambda_\omega g_\varrho^2 g_\omega^2 \sigma^2,
  \end{align}
  where $t_3 = \pm 1/2$ is the isospin.
  The parameters $g_\sigma$, $g_\omega$, $\rho_\varrho$, $b$, $c$, $\xi$ and $\Lambda_\omega$
  are sampled from Bayesian analysis.

  To obtain the EOS in NSs conditions,
  we need to impose beta equilibrium and charge neutrality:
  \begin{align}
    \mu_p &= \mu_n -\mu_e, \\
    0 &= \sum_{i = p,e,\mu} q_i \rho_i,
  \end{align}
  where $q_i$ is the electric charge.
  However, this model should also satisfy the nuclear matter properties,
  constrained by Bayesian inference (see Sec. \ref{sec:Bayesian}).
  Symmetric nuclear matter (SNM) and pure neutron matter (PNM) equations
  are solved by imposing $\rho_p = \rho_n$ and $\rho_p = 0$, respectively.

\subsection{Quark phase}
  We used two different models to describe the quark phase:
  the Nambu-Jona-Lasinio (NJL) model \cite{NJL1, NJL2} and the Mean Field Theory of QCD (MFTQCD) \cite{Fogaca:2010mf}.
  The NJL model is a widely used model that includes all global QCD symmetries
  and reproduces chiral symmetry breaking in the vacuum.
  The second model is obtained by making approximations in the gluon field
  and exhibits behavior similar to that of the vectorial MIT bag model.
  For both models, we imposed chemical equilibrium and charge neutrality.
  A brief description of each model is provided below.

\subsubsection{NJL}
  The Nambu-Jona-Lasinio (NJL) model \cite{NJL1, NJL2}
  is an effective model of point-like quark interactions.
  Despite the absence of gluons and a color confinement mechanism in this model,
  the NJL is well-suited for the description of large-density physics.
  This is due to its capacity to be designed to satisfy
  all the global symmetries of quantum chromodynamics (QCD) and
  to study manifestations of spontaneous chiral symmetry breaking
  \cite{VOGL1991195}.
  In this work, the SU(3) NJL Lagrangian is given by
  \begin{align}
    \mathcal{L} &= \bar{\psi} (i \gamma_\mu \partial^\mu -m +\mu \gamma^0) \psi  \nonumber \\
      &\quad +\frac{G}{2}
        \left[
          (\bar{\psi} \lambda_a \psi)^2
          +(\bar{\psi} i \gamma^5 \lambda_a \psi)^2
        \right] \nonumber \\
      &\quad +\kappa
        \left\{
          \det_f[\bar{\psi} (1 +\gamma^5) \psi]
          +\det_f[\bar{\psi} (1 -\gamma^5) \psi]
        \right\} \nonumber \\
      &\quad +\mathcal{L}_\text{int},
      \label{eq:NJL_L}
  \end{align}
  where $m$ and $\mu$ are the quark current masses and chemical potential matrices,
  $\lambda_a$, with $a = 1, 2, \dots, 8$, are the Gell-Mann matrices, and
  $\lambda_0$ is defined as $\lambda_0 = \sqrt{2/3} \mathbb{1}$.
  The second term of Eq. \ref{eq:NJL_L} is the standard NJL term
  responsible for chiral symmetry breaking in the vacuum.
  The third one is implemented to explicitly break the $U(1)_A$ symmetry,
  as this is not a vacuum symmetry in QCD.
  The last term represents quark interaction terms
  added to better describe the physics of NSs.
  Here, we consider the following terms:
  \begin{align}
    \mathcal{L}_\text{int}
      &= -G_\omega
        \left[
          (\bar{\psi} \gamma^\mu \lambda_0 \psi)^2
          +(\bar{\psi} \gamma^\mu \gamma^5 \lambda_0 \psi)^2
        \right] \nonumber \\
      &\quad -G_\rho
        \sum_{a=1}^8
        \left[
          (\bar{\psi} \gamma^\mu \lambda_a \psi)^2
          +(\bar{\psi} \gamma^\mu \gamma^5 \lambda_a \psi)^2
        \right] \nonumber \\
      &\quad -G_{\omega\omega}
        \left[
          (\bar{\psi} \gamma^\mu \lambda_0 \psi)^2
          +(\bar{\psi} \gamma^\mu \gamma^5 \lambda_0 \psi)^2
        \right]^2 \nonumber \\
      &\quad -G_{\sigma\omega}
        \sum_{a=0}^8
        \left[
          (\bar{\psi} \lambda_a \psi)^2
          +(\bar{\psi} i \gamma^\mu \gamma^5 \lambda_a \psi)^2
        \right]
        \nonumber \\ &\quad \times
        \left[
          (\bar{\psi} \gamma^\mu \lambda_0 \psi)^2
          +(\bar{\psi} \gamma^\mu \gamma^5 \lambda_0 \psi)^2
        \right] \nonumber \\
      &\quad -G_{\rho\omega}
        \sum_{a=1}^8
        \left[
          (\bar{\psi} \gamma^\mu \lambda_a \psi)^2
          +(\bar{\psi} \gamma^\mu \gamma^5 \lambda_a \psi)^2
        \right]
        \nonumber \\ &\quad \times
        \left[
          (\bar{\psi} \gamma^\mu \lambda_0 \psi)^2
          +(\bar{\psi} \gamma^\mu \gamma^5 \lambda_0 \psi)^2
        \right].
        \label{eq:NJL_int}
  \end{align}

  To obtain the EOS, we apply the mean-field approximation.
  Effective mass and chemical potential are given by the gap equations
  \begin{align}
    \tilde{m}_i &= m_i + -2 G \sigma_i -2 \kappa \sigma_j \sigma_k \nonumber \\
      &\quad
      +\frac{8}{3} G_{\sigma\omega} (\sigma_i^2 +\sigma_j^2 +\sigma_k^2) \sigma_i \\
    \tilde{\mu}_i &= \mu_i -
      \frac{4}{3} G_\omega (\rho_i +\rho_j +\rho_k)
      -\frac{4}{3} G_\rho (2 \rho_i -\rho_j -\rho_k) \nonumber \\
      &\quad
      -\frac{16}{9} G_{\omega\omega} (\rho_i +\rho_j +\rho_k)^3 \nonumber \\
      &\quad
      -\frac{8}{3} G_{\sigma\omega} (\sigma_i^2 +\sigma_j^2 +\sigma_k^2)
        (\rho_i +\rho_j +\rho_k) \nonumber \\
      &\quad
      -\frac{8}{9} G_{\rho\omega} (\rho_i +\rho_j +\rho_k) \nonumber \\
      &\quad
      \times (4 \rho_i^2 +\rho_j^2 +\rho_k^2 -\rho_i \rho_j -\rho_i \rho_k -4 \rho_j \rho_k),
  \end{align}
  with $i \neq j \neq k \in \{u, d, s\}$.
  For $T = 0$, the grand canonical potential is given by
  \begin{align}
    \Omega
      &= \Omega_0
        +G
        \left(
          \sigma_u^2 + \sigma_d^2 + \sigma_s^2
        \right)
        +4 \kappa \sigma_u \sigma_d \sigma_s \nonumber \\
      &\quad
        -\frac{2}{3} G_\omega (\rho_u + \rho_d + \rho_s)^2 \nonumber \\
      &\quad
        -\frac{4}{3} G_\rho
        (\rho_u^2 +\rho_s^2 +\rho_s^2 -\rho_u \rho_d -\rho_u \rho_s -\rho_d \rho_s)
      \nonumber \\ &\quad
        -\frac{4}{3} G_{\omega\omega} (\rho_u + \rho_d + \rho_s)^4 \nonumber \\
      &\quad
        -4 G_{\sigma\omega}
        \left( \sigma_u^2 + \sigma_d^2 + \sigma_s^2 \right)
        \left( \rho_u + \rho_d + \rho_s \right)^2
      \nonumber \\
      &\quad -\frac{8}{3} G_{\rho\omega}
        \left( \rho_u + \rho_d + \rho_s \right)^2 \nonumber \\
      &\qquad \times
        \left(
          \rho_u^2 + \rho_d^2 + \rho_s^2
          -\rho_u \rho_d -\rho_u \rho_s -\rho_d \rho_s
        \right) \nonumber \\
      &\quad -\frac{3}{\pi^2}
        \sum_{f=u,d,s} \int_{k_{F_f}}^\Lambda dp p^2 E_f
        -\frac{1}{\pi^2} \sum_{f=u,d,s} \tilde{\mu}_f k_{F_f},
  \end{align}
  where $\Omega_0$ is set to vanish the potential in the vacuum
  and $k_{F,f} = \sqrt{\tilde{\mu}_f^2 + \tilde{m}_f^2}$ is the Fermi momentum.
  Here, we used the 3-momentum cutoff scheme ($\Lambda$).
  Imposing that the grand canonical potential must be stationary with respect to
  $\sigma_i$ and $\rho_i$ \cite{buballa}, i.e.,
  \begin{equation}
    \frac{\partial \Omega}{\partial \sigma_i} =
    \frac{\partial \Omega}{\partial \rho_i} =
    0,
  \end{equation}
  we obtain
  \begin{align}
    \sigma_f &=
      -\frac{3}{\pi^2}
      \int_{k_{F_f}}^\Lambda dp p^2 \frac{\tilde{m}_f}{E_f}, \\
    \rho_f &= \frac{1}{\pi^2} k_{F_f}^3,
  \end{align}
  at zero temperature.

  The parameters $G$, $\kappa$, $m_i$ and $\Lambda$ are set to satisfy the
  mass and decay constant experimental data from $\pi^\pm$, $K^\pm$, $\eta$ and $\eta'$
  \cite{Olive_2014} (see Table \ref{tab:fixed_njl}).
  The coupling constants
  $G_\omega$, $G_\rho$, $G_{\omega\omega}$, $G_{\sigma\omega}$ and $G_{\rho\omega}$
  are sampled from Bayesian analysis.

  \begin{table}[ht]
    \centering
    \setlength{\tabcolsep}{7.pt}
    \renewcommand{\arraystretch}{1.2}
    \begin{tabular}{ccccc}
      \hline \hline
      $\Lambda$ (MeV) & $m_{u,d}$ (MeV) & $m_s$ (MeV) & $G \Lambda^2$ & $\kappa \Lambda^5$ \\
      \hline
      623.58 & 5.70 & 136.60 & 3.34 & -13.67 \\
      \hline
    \end{tabular}
    \caption{Fixed parameters for NJL model,
      set to satisfy the experimental data from \cite{Olive_2014}.}
    \label{tab:fixed_njl}
  \end{table}

  Furthermore, we set $P \to P +B$, where $B$ is a constant.
  This parameter exhibits behavior similar to the bag constant in the MIT bag model,
  strongly influencing the location of the phase transition point.
  This parameter is also sampled from Bayesian inference. 
  The effects of the different 
   multiquark interaction channels of Eq.~(\ref{eq:NJL_int}) on the properties of hybrid stars, namely the interplay between the eight-quark vector interaction and the four-quark isovector-vector interaction, as well as higher-order repulsive interactions, have been studied in \cite{Ferreira:2020evu,Ferreira:2020kvu,Ferreira:2021osk}.

\subsubsection{MFTQCD}
  In the mean field theory of quantum chromodynamics (MFTQCD),
  a decomposition of the gluon fields into low (soft) and high (hard) gluons
  is assumed in the QCD Lagrangian
  \cite{Fogaca:2010mf, Franzon:2012in, Albino:2021zml},
  i.e.,
  \begin{equation}
    \tilde{G}^{a\mu}(k) = \tilde{A}^{a\mu}(k) +\tilde{\alpha}^{a\mu}(k),
  \end{equation}
  where $G$ is the gluon field in  momentum space and
  $A$ and $\alpha$ are the soft and hard gluon fields, respectively.
  Due to their small momenta, soft gluons are approximately constant
  and are replaced by their expected values in vacuum, given by
  \cite{PhysRevD.34.1591, PhysRevD.71.074007}
  \begin{align}
    \left< A^{a\mu} A^{b\nu} \right>
        &= -\frac{\delta^{ab}}{8} \frac{g^{\mu\nu}}{4} \mu_0^2, \\
    \left< A^{a\mu} A^{b\nu} A^{c\rho} A^{d\eta} \right>
        &= \frac{\phi_0^4}{(32)(34)}
        \left[
            g_{\mu\nu} g^{\rho\eta} \delta^{ab} \delta^{cd} 
        \right. \nonumber \\
        &\quad \left. 
            +g_\mu^\rho g_\nu^\eta \delta^{ac} \delta^{bd} 
            +g_\mu^\eta g_\nu^\rho \delta^{ad} \delta^{bc} 
        \right],
  \end{align}
  where $\phi_0$ and $\mu_0$ are energy scales to be determined.
  Assuming hard gluons have a large occupation number at all energy levels,
  they can be replaced by classical fields
  \cite{Serot:1984ey}
  \begin{equation}
    \alpha_\mu^a \to \left< \alpha_\mu^a \right> = \alpha_0^a \delta_{\mu 0},
  \end{equation}
  where $\alpha_0$ is a constant.
  The MFTQCD Lagrangian is obtained after a straightforward calculation
  \begin{align}
    \mathcal{L}_\text{MFTQCD} &=
        -B
        +\frac{m_G^2}{2} \alpha_0^a \alpha_0^a
        \nonumber \\ &\quad
    +\sum_{q=1}^{N_f} \bar{\psi}_i^q
    \left( 
      i \delta_{ij} \gamma^\mu \partial_\mu
      + g \gamma^0 T_{ij}^a \alpha_0^1
      -\delta_{ij} m_q
    \right) \psi_j^q, 
  \end{align}
  where
  \begin{align}
    m_G^2 &= \frac{9}{32} g^2 \mu_0^2, \\
    B &= \frac{9}{4(34)} g^2 \phi_0^4 = \left< \frac{1}{4} F^{a\mu\nu} F^a_{\mu\nu} \right>.
  \end{align}
  These are defined due to the fact that $m_G$ acts as
  a gluon mass in the MFTQCD Lagrangian
  and that $B$ exhibits behavior similar to the MIT bag constant.
  Using the energy-momentum tensor to calculate the equations of motion yields the EOS
  \begin{align}
    P &=  \frac{27}{2} \xi^2 \rho_B^2 -B +P_F, \\
    \epsilon &= \frac{27}{2} \xi^2 \rho_B^2 +B +\epsilon_F,
      \label{eq:mqcd}
  \end{align}
  where $P_F$ and $\epsilon_F$ are the pressure and energy density
  of a noninteracting Fermi gas of quarks and electrons, and
  $\xi \equiv g/m_G$.
  A more detailed deduction can be found in \cite{Fogaca:2010mf}.
  The final EOS exhibits behavior similar to that of a vectorial MIT bag model,
  in which the term $\xi$ acts as the vectorial term.
  In this work, the following values of masses were used:
  $m_u = 5 \text{ MeV}$, $m_d = 7 \text{ MeV}$ and $m_s = 100 \text{ MeV}$.
  The values of $\xi$ and $B$ are sampled by Bayesian inference. Note that recommended values from the Particle Data Group \cite{ParticleDataGroup:2022pth} for the up- and down-quark masses ($m_u = 2.16$~MeV and $m_d = 4.70$~MeV) were not  considered, but the results are insensitive to these values.

\section{Bayesian Approach} \label{sec:Bayesian}
  Bayesian analysis samples the parameters of a model to satisfy the constraints which were imposed.
  This process is performed using Bayes' theorem, given by
  \begin{equation}
    P(\text{model}|\text{data}) =
        \frac{P(\text{data}|\text{model}) P(\text{model})}{P(\text{data})},
  \end{equation}
  where $P(\text{model}|\text{data})$ is the posterior distribution,
  $P(\text{data}|\text{model})$ the likelihood,
  $P(\text{model})$ the prior, and
  $P(\text{data})$ the evidence.
  In this work, we use the PyMultiNest \cite{pymultinest,pymultinest2} sampler,
  which is based on the nested sampling method
  and part of the Bayesian inference library BILBY \cite{Ashton:2018jfp}.
  To apply this method, we must define the prior and likelihood distributions.
  The prior probability is the initial distribution of the parameters.
  Here, we use the uniform distribution defined in
  Table \ref{tab:prior_Hyb} (hybrid sets) and \ref{tab:prior_RMF} (hadron set).
  We discuss five datasets: the hadronic set RMF and the hybrid sets NJL, NJL-GW, r-NJL and MFTQCD.
  The lowest and highest values of these distributions 
  {were chosen such that the posterior distributions did not show unjustified restrictions,  except for  the set r-NJL, for which the hadronic parameters were considered the same as the ones taken for the RMF set. This allows us to discuss, when building the hybrid EOS, the effect of forcing the hadronic prior space to coincide with the one considered for the hadronic EOS. Quark matter in sets NJL, NJL-GW, r-NJL is described by the NJL model and  the quark model MFTQCD is used in the set MFTQCD.}
  
  To determine the likelihood
  -- that is, the probability of obtaining the restrictions imposed for a particular parameter set --
  we considered the nuclear matter properties (NMP) and pure neutron matter (PNM) data,
  presented in Table \ref{tab:nmp_constraints},
  and the X-ray NICER Data from
  J0030+0451 \cite{Vinciguerra_2024,Miller_2019},
  J0740+6630 \cite{Salmi_2024,Miller:2021qha} and
  J0437+4715 \cite{Choudhury_2024,Reardon_2024}.
  We also ensured causality within NSs and
  a pQCD constraint at $7 \rho_0$ developed in \cite{Gorda_2023, komoltsev_2023_7781233}.
  For the NJL-GW set, we imposed gravitational wave (GW) data from GW170817 \cite{Abbott_2018}.
  For the hybrid sets, we imposed constraints to ensure a phase transition.
  All details are in Appendix \ref{sec:Bayes_Details}.

	The evidence, which is only a normalization term,
	is estimated by the nested sampling method.
	We used 1,000 live points in all sets of equations
	and obtained 6514, 4879, 6037, 7521, and {5327} samples for the
	NJL, MFTQCD, RMF, NJL-GW and {r-NJL} sets, respectively.

\begin{table*}[ht]
    \centering
    \setlength{\tabcolsep}{12.pt}
    \renewcommand{\arraystretch}{1.2}
    \begin{tabular}{ccccccc}
        \hline \hline
        \multicolumn{7}{c}{Set NJL}                                                  \\
        \multicolumn{3}{c}{NJL}             &  & \multicolumn{3}{c}{RMF}             \\ 
        \hline 
        Parameters              & min & max &  & Parameters       & min    & max     \\
        \hline
        $\xi_\omega$            & 0   & 0.5 &  & $g_\sigma$       &  9     &  12     \\
        $\xi_\rho$              & 0   & 1   &  & $g_\omega$       &  11    &  15     \\
        $\xi_{\omega\omega}$    & 0   & 30  &  & $g_\rho$         &  9.546 &  15.000 \\
        $\xi_{\sigma\omega}$    & 0   & 8   &  & BB                &  1.500 &  3.500  \\
        $\xi_{\rho\omega}$      & 0   & 50  &  & CC                & -4.627 & -1.500  \\
        $B$ (MeV/fm$^3$)        & 0   & 30  &  & $\xi$            &  0     &  0.016  \\
                                &     &     &  & $\Lambda_\omega$ &  0     &  0.103  \\
        \hline \hline
        \multicolumn{7}{c}{Set MFTQCD}                                                  \\
        \multicolumn{3}{c}{MFTQCD}             &  & \multicolumn{3}{c}{RMF}             \\ 
        \hline 
        Parameters              & min & max    &  & Parameters       & min    & max     \\
        \hline
        $\xi_Q$ (MeV$^{-1}$)    & 0   & 0.0018 &  & $g_\sigma$       &  7     &  10     \\
        $B$ (MeV/fm$^3$)        & 50  & 180    &  & $g_\omega$       &  8     &  13     \\
                                &     &        &  & $g_\rho$         &  8.000 &  15.000 \\
                                &     &        &  & BB                &  1.000 &  9.000  \\
                                &     &        &  & CC                & -5.000 &  5.000  \\
                                &     &        &  & $\xi$            &  0     &  0.040  \\
                                &     &        &  & $\Lambda_\omega$ &  0     &  0.120  \\
        \hline \hline 
        \multicolumn{7}{c}{Set NJL-GW}                                              \\
        \multicolumn{3}{c}{NJL}             &  & \multicolumn{3}{c}{RMF}            \\ 
        \hline 
        Parameters              & min & max &  & Parameters       & min    & max    \\
        \hline
        $\xi_\omega$            & 0   & 0.5 &  & $g_\sigma$       &  8     & 11     \\
        $\xi_\rho$              & 0   & 1   &  & $g_\omega$       & 10     & 14     \\
        $\xi_{\omega\omega}$    & 0   & 30  &  & $g_\rho$         &  9.546 & 15     \\
        $\xi_{\sigma\omega}$    & 0   & 8   &  & BB                &  1.500 &  3.500 \\
        $\xi_{\rho\omega}$      & 0   & 50  &  & CC                & -4.627 & -1.500 \\
        $B$ (MeV/fm$^3$)        & 0   & 30  &  & $\xi$            &  0     & 0.016 \\
                                &     &     &  & $\Lambda_\omega$ &  0     & 0.103 \\
        \hline \hline 
        \multicolumn{7}{c}{Set r-NL}                                                  \\
        \multicolumn{3}{c}{NJL}             &  & \multicolumn{3}{c}{RMF}            \\ 
        \hline 
        Parameters              & min & max &  & Parameters       & min    & max    \\
        \hline
        $\xi_\omega$            & 0   & 0.5 &  & $g_\sigma$       &  8.010 & 9.691  \\
        $\xi_\rho$              & 0   & 1   &  & $g_\omega$       &  9.084 & 12.167 \\
        $\xi_{\omega\omega}$    & 0   & 30  &  & $g_\rho$         &  9.546 & 14.599 \\
        $\xi_{\sigma\omega}$    & 0   & 8   &  & BB                &  2.205 & 6.903  \\
        $\xi_{\rho\omega}$      & 0   & 50  &  & CC                & -4.627 & 3.530  \\
        $B$ (MeV/fm$^3$)        & 0   & 30  &  & $\xi$            &  0     & 0.016  \\
                                &     &     &  & $\Lambda_\omega$ &  0.036 & 0.103  \\
        \hline
    \end{tabular}
    \caption{The lowest and highest values for the uniform distribution prior used for the data sets NJL, NJL-GW, MFTQCD and r-NJL.
    We defined $BB = b \times 10^3$ and $CC = c \times 10^3$.}
    \label{tab:prior_Hyb}
\end{table*}

\begin{table}[ht]
    \centering
    \setlength{\tabcolsep}{12.pt}
    \renewcommand{\arraystretch}{1.2}
    \begin{tabular}{ccc}
        \hline \hline
        \multicolumn{3}{c}{Set RMF}        \\
        \hline 
        Parameters       & min    & max    \\
        \hline 
        $g_\sigma$       &  6.5   & 13     \\
        $g_\omega$       &  6.5   & 15.5   \\
        $g_\rho$         &  6.5   & 16.5   \\
        BB                &  0.500 &  9.000 \\
        CC                & -5.000 &  5.000 \\
        $\xi$            &  0     &  0.040 \\
        $\Lambda_\omega$ &  0     &  0.120 \\
        \hline
    \end{tabular}
    \caption{The lowest and highest values for the uniform distribution prior used for the data set RMF.
    We defined $BB = b \times 10^3$ and $CC = c \times 10^3$.}
    \label{tab:prior_RMF}
\end{table}

\begin{table*}[ht]
    \centering
    \setlength{\tabcolsep}{12.pt}
    \renewcommand{\arraystretch}{1.2}
    \begin{tabular}{ccccccc}
        \hline \hline 
        \multicolumn{3}{c}{NMP}              &  & \multicolumn{3}{c}{PNM}                    \\ 
        Quantity & value/band & Ref.         & & Quantity & value/band & Refs. \\
        \hline
        $\frac{dEA}{d\rho}$ & 0              &
        & &
        $EA_{PNM}(\rho = 0.05)$ & $6.8 \pm 1.02$ & \cite{Huth:2021bsp} \\
        $EA_0$              & $-16 \pm 0.2$ & \cite{PhysRevC.90.055203} 
        & & 
        $EA_{PNM}(\rho = 0.10)$ & $10.5 \pm 1.97$ & \cite{Huth:2021bsp} \\
        $K_0$               & $230 \pm 30$   & \cite{Todd-Rutel:2005yzo,Shlomo:2006ole}
        & &
        $EA_{PNM}(\rho = 0.15)$ & $15.3 \pm 3.44$ & \cite{Huth:2021bsp} \\
        $J_{sym,0}$         & $32.5 \pm 1.8$ & \cite{PhysRevC.104.065804}
        & &
        & \\
        \hline
    \end{tabular}
    \caption{Constraints for NMP and PNM.}
    \label{tab:nmp_constraints}
\end{table*}

\section{Results} \label{sec:Results}
  In this section, we present the EOSs,  as well as other properties such as the mass-radius diagram, speed of sound, and trace anomaly, calculated using samples from Bayesian inferences.

%
%
%

  \subsection{Properties of the EOS}
  
\begin{figure}[ht]
    \centering
    \includegraphics[width=\linewidth]{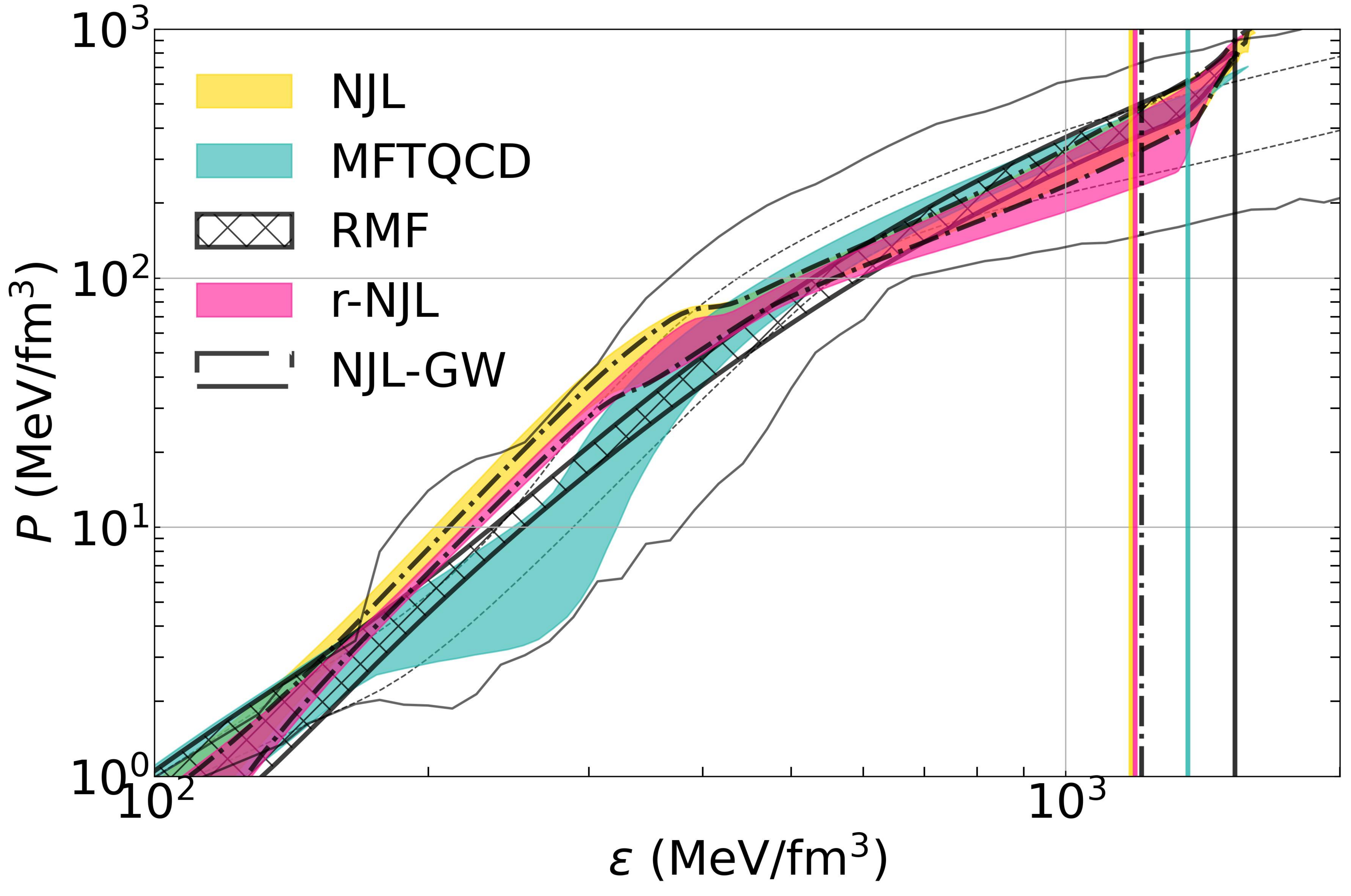}
    \caption{Pressure versus energy density of the 90\% of CI.
      Sets NJL, MFTQCD, RMF, r-NJL and NJL-GW
      are represented in yellow, cyan and hatched, pink and dot-dashed bands, respectively.
      Vertical lines indicate the 90\% CI maximum for $\epsilon$ at the maximum NS mass.
      Band in gray represents the full (solid) and 90\% of CI (dashed)
      of model-independent results from \cite{Altiparmak:2022bke}.}
    \label{fig:paper2_PxE}
  \end{figure}

  Sets NJL, MFTQCD, RMF, r-NJL and NJL-GW are represented in Fig. \ref{fig:paper2_PxE}
  where the 90\% credible intervals (CI) of pressure versus energy density for these sets are plotted
  in yellow, cyan and hatched, pink and dot-dashed bands, respectively.
  When we compare the hybrid sets, we see that the MFTQCD set
  allows for a phase transition at low energy densities.
  This can also be seen in Table \ref{tab:NS_prop},
  which shows some numerical results,
  including the phase transition density.
  MFTQCD allows a phase transition at a density of $\sim 0.170 \text{ fm}^{-3}$ (minimum 90\% of CI),
  very close to the saturation density.
  In contrast, the three NJL sets allow a phase transition at a density of
  approximately twice the saturation density.
  A similar result was obtained in \cite{Albino:2024ymc},
  where the hadron phase was described by a fixed RMF equation,
  and Bayesian inference was applied only to the quark phase parameters.
  
  By comparing the NJL model sets,
  it can be seen that the NJL-GW and r-NJL sets have lower pressure than the NJL set
  in the hadron phase region ($\epsilon \lesssim 400 \text{ MeV/fm}^3$).
  Therefore, applying the GW170817 constraint or
  restricting the hadron parameters results in a softer hadron phase,
  although the latter has a stronger effect.

  It is interesting to compare the region spanned by our models in the $P-\epsilon$ space with those determined in \cite{Altiparmak:2022bke},  using a model-independent  method based on the  sound speed as a function of the chemical potential introduced in \cite{Annala_2020}.
  
  All sets are compatible with the full model-independent results from \cite{Altiparmak:2022bke};
  see the gray border in Fig. \ref{fig:paper2_PxE}.
  Note, however, that the envelope-region defined in \cite{Altiparmak:2022bke} was also constrained by the GW170817 detection, a condition that we have only imposed to the  NJL-GW set.
  In addition, the $\chi$EFT constraint we have applied in our analysis,   the energy per neutron given in \cite{Huth:2021bsp}, is slightly different from that  considered in \cite{Altiparmak:2022bke}, 
  the NS matter pressure given in \cite{Hebeler_2013}.
  This explains why the NJL distribution may spread outside the envelope defined in \cite{Altiparmak:2022bke}.
  Note that the RMF model follows the 90\%CI band of \cite{Altiparmak:2022bke}, approximately.
  The MFTQCD model spans the 90\%CI band and also covers a range below this band
  with the EOS already in the quark phase. The population of this low energy density region requires a phase transition to quark matter.
  The three NJL model sets have a very hard hadronic EOS for densities beyond the $\chi$EFT band,
  especially the NJL set.
  This region extends above the 90\% CI band until the transition to the quark phase.
  The quark phase is compatible with the 90\%CI band of \cite{Altiparmak:2022bke}.
  Large quark cores are possible for particularly stiff hadronic EOSs.

  \subsection{Speed of Sound}
  
  \begin{figure}[ht]
    \centering
    \includegraphics[width=\linewidth]{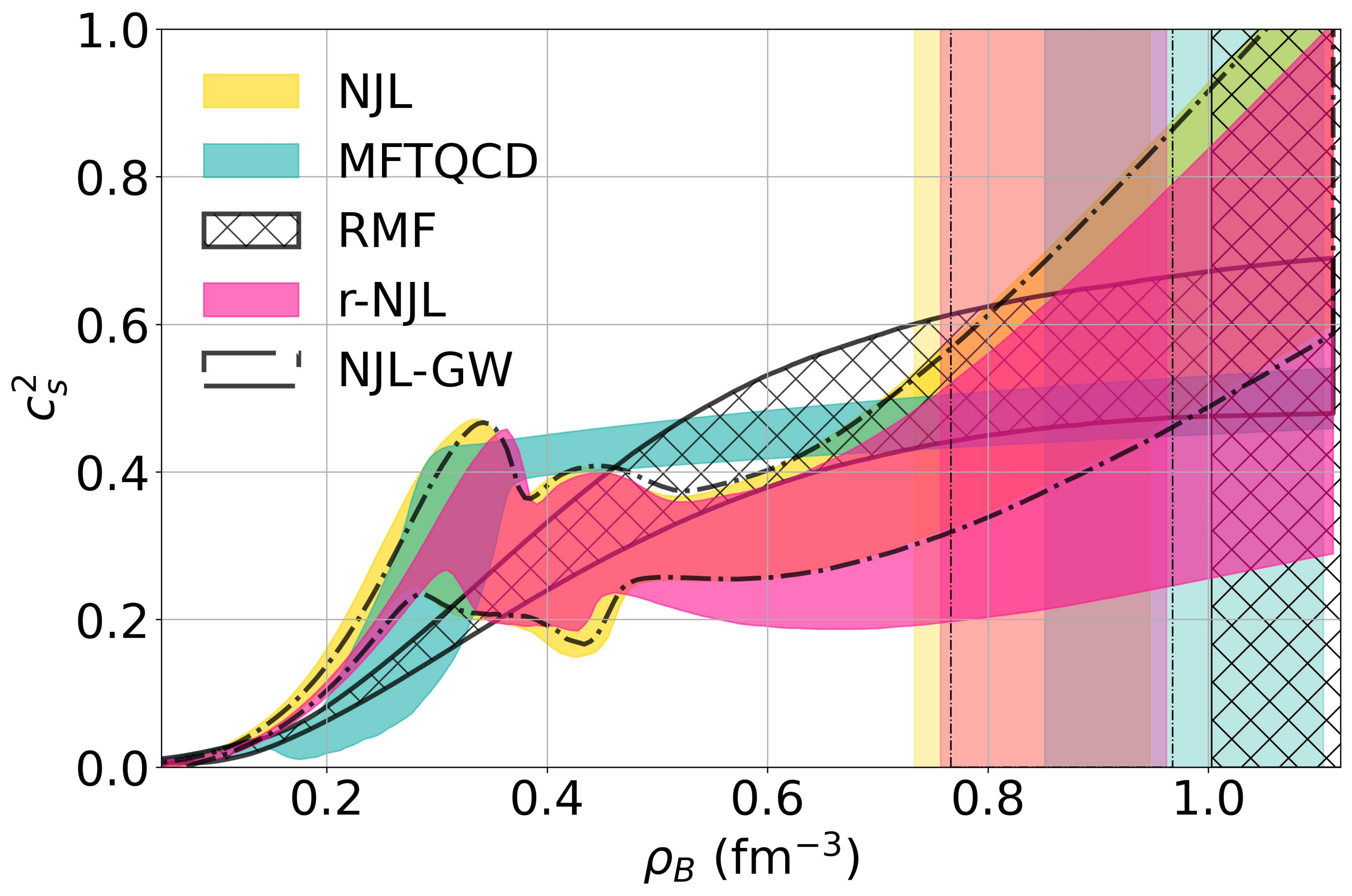}
    \caption{
      Speed of sound {squared in units of $c^2$} 90\% CI distributions versus the  baryonic density.
      Same color code from Fig. \ref{fig:paper2_PxE}.
      Vertical bands represent the central density at the maximum NS mass of the 90\% CI.}
    \label{fig:paper2_cs2xrho}
  \end{figure}

  Fig. \ref{fig:paper2_cs2xrho} shows the speed of sound squared versus baryonic density.
  The vertical bands represent the density at maximum NS mass.
  Interestingly, the speed of sound for each set is significantly different.
  For the NJL, NJL-GW and r-NJL sets, there are two bumps: the first is caused by the phase transition,
  and the second is caused by the appearance of the strange quark.
  The phase transition bump occurs at slightly lower densities for the NJL set
  and at larger values for the r-NJL.
  The NJL hadronic phase is stiffer and therefore the phase transition density to quark matter is
  lower for NJL than for NJL-GW and r-NJL sets (see Table \ref{tab:NS_prop}).
  However, this difference is less noticeable for the second bump,
  and results essentially due to the small differences of the couplings associated to the flavor dependent terms.
  After that, the speed of sound increases with density
  due to the term $\xi_{\omega\omega}$ \cite{Ferreira:2021osk,Albino:2024ymc}.
  However, the EOSs are causal at the central density of maximum mass configurations
  (see the vertical band).
  The speed of sound squared takes values on the order of 0.4-0.6 for all models, at the center of maximum mass stars.
  The MFTQCD and RMF sets have similar speeds of sound at large densities,
  even though they have different components (quark and hadron respectively).
  They both increase until $\rho_B \approx 0.3 \text{ fm}^{-3}$ (MFTQCD)
  and $\rho_B \approx 0.6 \text{ fm}^{-3}$ (RMF),
  then stabilize at $c_s^2 = 0.5$.
  
  As shown by the vertical bands, the NJL, NJL-GW, and r-NJL sets have smaller central densities for maximum-mass stars, which can reach values as high as  $\sim 0.95\text{ fm}^{-3}$ at 90\%CI.
The MFTQCD set can reach larger values central densities, above  $1.10\text{ fm}^{-3}$ at 90\%CI.
  See Table \ref{tab:NS_prop} for more numerical details.

  \subsection{Mass-Radius Diagram Result}
\begin{figure}[ht]
    \centering
    \includegraphics[width=.9\linewidth]{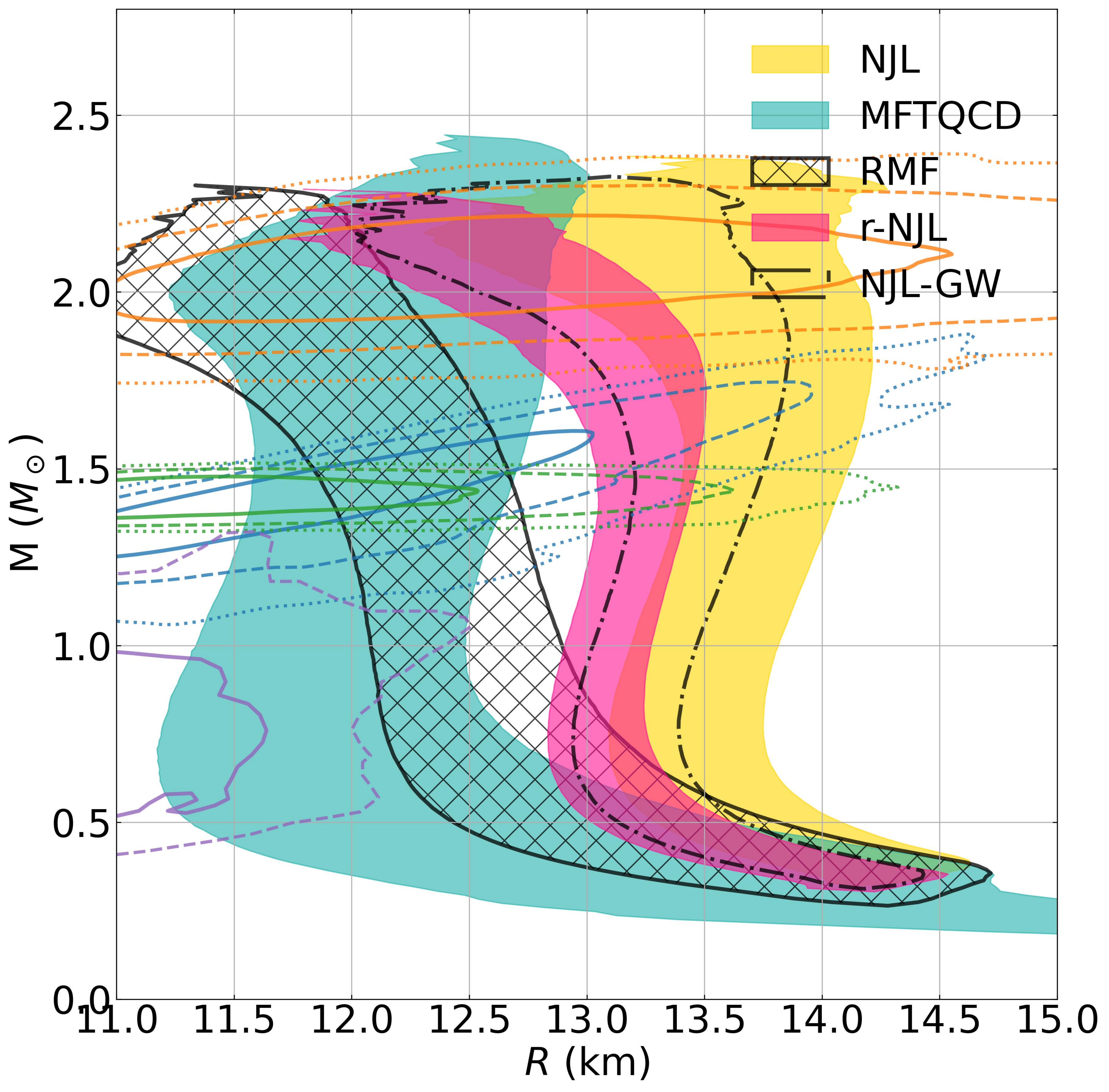}
    \caption{Mass-radius diagram of the 90\% of CI.
      Same color code from Fig.  \ref{fig:paper2_PxE}.
      Observational data are shown as
      PSR J0030+0451 (blue) \cite{Vinciguerra_2024,Miller_2019},
      PSR J0740+6620 (orange) \cite{Salmi_2024,Miller:2021qha},
      PSR J0437+4715 (green) \cite{Choudhury_2024} and
      HESS J1731-347 (purple) \cite{hess}
      with $1 \sigma$, $2 \sigma$ and $3 \sigma$ represented by
      solid, dashed and dotted lines, respectively.}
    \label{fig:paper2_MxR}
  \end{figure}

  The mass-radius diagram shown in Fig. \ref{fig:paper2_MxR}
  was obtained by solving the Tolman-Oppenheimer-Volkov (TOV) equations \cite{Tolman:1939jz,Oppenheimer:1939ne}.
  The following observational data are represented in this figure
  with $1 \sigma$ (solid), $2 \sigma$ (dashed) and $3 \sigma$ (dotted):
  PSR J0030+0451 (blue) \cite{Vinciguerra_2024,Miller_2019},
  PSR J0740+6620 (orange) \cite{Salmi_2024,Miller:2021qha}, and
  PSR J0437+4715 (green) \cite{Choudhury_2024,Reardon_2024}
  by NICER, and
  HESS J1731-347 (purple) \cite{hess}.
  Comparing the hybrid sets with the hadron RMF set,
  we see that the NJL, NJL-GW and r-NJL sets shift the mass-radius diagram to the right,
  while the MFTQCD set shifts it to the left.
  Due to the small phase transition densities,
  the MFTQCD set can reach smaller radii than the other sets.
  It is the only set that is compatible with the HESS data (purple stain) at 68\%.
  This result seems to indicate that the onset of quarks could occur at low densities.
  As a consequence, low mass NS could be hybrid stars.
  As we will discuss later, the lower the transition density to quark matter
  (and, therefore, the lower $M_\text{trans}$),
  the more compatible the MFTQCD curves are with the HESS data (see Fig. \ref{fig:Mtrans}).
  
  As previously mentioned, the NJL model 
  yields larger radii than RMF EOS.
  This occurs because the NJL EOS is  stiffer than the MFTQCD EOS,
  and in order for the star to contain an appreciable amount of quark matter,
  the hadron phase should be described by a stiff EOS.
  To attain two solar mass stars, the term $\xi_{\omega\omega}$ plays an important role:
  at low density its contribution is small, but its importance increases with density,
  making the quark EOS stiff enough.
  A stiff hadron EOS implies an increase in the radius of low mass stars.
  Note that the vector terms of the quark EOS allow for quite large radii at $\sim 2 \text{M}_\odot$,
  compatible with the NICER results for J0740+6620.
  
   \begin{figure}[ht]
    \centering
    \includegraphics[width=.9\linewidth]{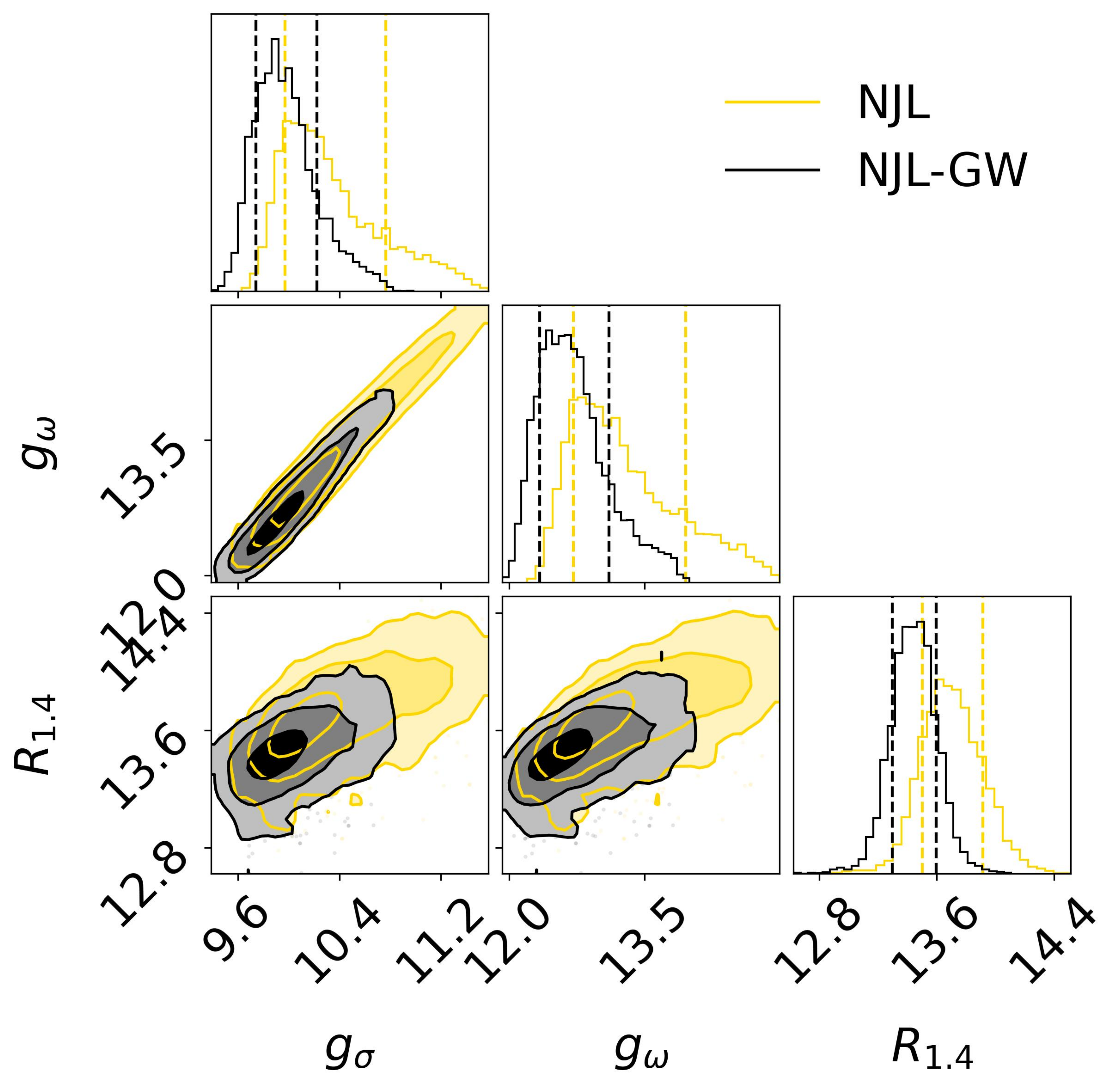}
    \caption{Corner plot of  $g_\sigma$, $g_\omega$ and $R_{1.4}$
      for the NJL and NJL-GW sets with different priors.}
    \label{fig:paper2_cornerH_prior}
  \end{figure}

  Comparing the NJL and NJL-GW sets,
  even though these hybrid sets use the same models for the hadron and quark phases,
  the mass-radius relations differ significantly.
  The NJL-GW sets have a notably smaller radius than the NJL set.
  The most significant difference is the maximum 90\% CI of $R_{1.4 \text{M}_\odot}$.
  The difference between the maximums is 0.353 km,
  while the difference between the minimums is $\sim$0.197 km
  (see Table \ref{tab:NS_prop}).
  This implies that the constraint imposed by GW170817
  restricts the radius from increasing too much.
  This difference is explained by the correlation between
  $g_\omega$ and $R_{1.4 \text{M}_\odot}$,
  as seen in Fig. \ref{fig:paper2_cornerH_prior}.
  Notably, $R_{1.4 \text{M}_\odot}$ increases with $g_\omega$ because a larger $g_{\omega}$ gives rise to a stiffer hadron EOS.
  For the NJL-GW set,  $g_{\omega}$ cannot reach
  as large values as for the NJL set.
  However, the minimum value of this coupling remains almost the same.
  The correlation  between couplings $g_{\sigma}$ and $g_{\omega}$ is imposed by the binding energy at saturation.
 
  Analyzing the r-NJL set, this set describes the smallest radii compared to the other NJL hybrid sets.
  However, it also has the smallest mass of all sets,
  with $M_\text{max}^\text{med} = 1.996 \text{M}_\odot$.
  Although the r-NJL set is mostly stiffer than the RMF set (see Fig. \ref{fig:paper2_PxE}),
  the fact that the RMF set is stiffer at high densities is sufficient for it to reach higher mass values.
%
  Except for set r-NJL, there is an increase in maximum mass for the hybrid models.
  This is possible due to the vector terms present in the quark models.
  However, all sets can describe NICER data.

    \begin{table}[hb]
    \centering
    \setlength{\tabcolsep}{12.pt}
    \renewcommand{\arraystretch}{1.2}
    \begin{tabular}{cccc}
      \hline \hline
      \multicolumn{4}{c}{Set RMF} \\
      quant & median & min & max \\
      \hline
      $M_\text{max}$           & 2.039    & 1.905    & 2.185    \\
      $R_\text{max}$           & 10.761   & 10.339   & 11.195   \\
      $\rho_\text{max}$        & 1.103    & 1.003    & 1.212    \\
      $\epsilon_\text{max}$    & 1354.565 & 1191.639 & 1533.405 \\
      $c_{s,\text{max}}^2$     & 0.582    & 0.478    & 0.692    \\
      $R_{1.4 \text{M}_\odot}$ & 12.297   & 11.923   & 12.714   \\
      \hline
    \end{tabular}
    \caption{The 90\% CI of the following quantities:
    {maximum mass ($M_\text{max}$, in M$_\odot$) and
    radius ($R_\text{max}$, in km),
    density ($\rho_\text{max}$, in fm$^{-3}$) and
    energy density ($\epsilon_\text{max}$, in MeV/fm$^3$),
    speed of sound ($c_{s,\text{max}}^2$) at the maximum NS mass;
    radius of the $1.4 \text{M}_\odot$ ($R_{1.4 \text{M}_\odot}$, in km).
    Results of the hadronic set.
    }}
    \label{tab:NS_prop_RMF}
  \end{table}

  \begin{table*}[ht]
    \centering
    \setlength{\tabcolsep}{12.pt}
    \renewcommand{\arraystretch}{1.2}
    \begin{tabular}{ccccccc}
      \hline \hline
      & \multicolumn{3}{c}{Set NJL} & \multicolumn{3}{c}{Set NJL-GW} \\
      quant & median & min & max & median & min & max \\
      \hline
      $\rho_\text{trans}$            & 0.353   & 0.304   & 0.388    & 0.362    & 0.314   & 0.391    \\
      $P_\text{trans}$               & 57.466  & 35.143  & 78.623   & 55.989   & 33.814  & 77.024   \\
      $\epsilon_\text{H,trans}$      & 350.549 & 295.526 & 391.007  & 358.957  & 304.433 & 393.36   \\
      $\epsilon_\text{Q,trans}$      & 395.697 & 331.009 & 468.817  & 395.183  & 334.473 & 458.06   \\
      $\Delta \epsilon_\text{trans}$ & 44.971  & 22.492  & 91.833   & 36.207   & 19.691  & 76.780   \\
      $M_\text{max}$                 & 2.130   & 2.018   & 2.236    & 2.108    & 1.993   & 2.212    \\
      $R_\text{max}$                 & 12.466  & 11.716  & 13.122   & 12.147   & 11.526  & 12.827   \\
      $\rho_\text{max}$              & 0.829   & 0.733   & 0.947    & 0.871    & 0.767   & 0.968    \\
      $\epsilon_\text{max}$          & 995.402 & 857.809 & 1178.842 & 1053.422 & 898.290 & 1211.631 \\
      $c_{s,\text{max}}^2$           & 0.529   & 0.329   & 0.776    & 0.564    & 0.340   & 0.795    \\
      $M_\text{Q,max}$               & 1.103   & 0.757   & 1.511    & 1.176    & 0.777   & 1.555    \\
      $R_\text{Q,max}$               & 7.955   & 6.956   & 8.933    & 8.069    & 7.026   & 8.952    \\
      $R_{1.4 \text{M}_\odot}$       & 13.695  & 13.390  & 14.051   & 13.448   & 13.193  & 13.698   \\
      $M_\text{trans}$               & 1.728   & 1.341   & 2.008    & 1.634    & 1.253   & 1.940    \\
      \hline \hline
      & \multicolumn{3}{c}{Set MFTQCD} & \multicolumn{3}{c}{Set r-NJL} \\
      quant & median & min & max & median & min & max \\
      \hline
      $\rho_\text{trans}$            & 0.222    & 0.170    & 0.308    & 0.369    & 0.333   & 0.395    \\
      $P_\text{trans}$               & 5.752    & 2.451    & 15.851   & 54.239   & 36.589  & 70.712   \\
      $\epsilon_\text{H,trans}$      & 208.795  & 157.329  & 295.494  & 364.841  & 324.201 & 395.105  \\
      $\epsilon_\text{Q,trans}$      & 289.862  & 254.155  & 337.472  & 399.421  & 349.543 & 446.832  \\
      $\Delta \epsilon_\text{trans}$ & 77.373   & 24.691   & 122.923  & 33.315   & 20.009  & 60.217   \\
      $M_\text{max}$                 & 2.133    & 1.970    & 2.315    & 1.996    & 1.863   & 2.154    \\
      $R_\text{max}$                 & 11.273   & 10.685   & 11.954   & 12.199   & 11.547  & 12.847   \\
      $\rho_\text{max}$              & 0.976    & 0.852    & 1.104    & 0.861    & 0.757   & 0.962    \\
      $\epsilon_\text{max}$          & 1205.841 & 1052.149 & 1361.615 & 1013.698 & 863.204 & 1191.993 \\
      $c_{s,\text{max}}^2$           & 0.487    & 0.458    & 0.515    & 0.388    & 0.200   & 0.744    \\
      $M_\text{Q,max}$               & 2.018    & 1.781    & 2.224    & 1.010    & 0.572   & 1.432    \\
      $R_\text{Q,max}$               & 10.266   & 9.405    & 11.101   & 7.818    & 6.471   & 8.633    \\
      $R_{1.4 \text{M}_\odot}$       & 12.111   & 11.571   & 12.649   & 13.265   & 13.050  & 13.436   \\
      $M_\text{trans}$               & 0.327    & 0.174    & 0.630    & 1.547    & 1.248   & 1.771    \\
      \hline
    \end{tabular}
    \caption{The 90\% CI of the following quantities:
    {the phase transition
    density ($\rho_\text{trans}$, in fm$^{-3}$),
    pressure ($P_\text{trans}$, in MeV/fm$^3$),
    hadronic energy density ($\epsilon_\text{H,trans}$, in MeV/fm$^3$) and
    quark energy ensity ($\epsilon_\text{Q,trans}$, in MeV/fm$^3$); 
    measure of the strength of the phase transition
    ($\Delta \epsilon_\text{trans} = \epsilon_\text{Q,trans} -\epsilon_\text{H,trans}$, in MeV/fm$^3$);
    maximum mass ($M_\text{max}$, in M$_\odot$) and
    radius ($R_\text{max}$, in km),
    density ($\rho_\text{max}$, in fm$^{-3}$),
    energy density ($\epsilon_\text{max}$, in MeV/fm$^3$),
    speed of sound ($c_{s,\text{max}}^2$)
    quark core mass ($M_\text{Q,max}$, in M$_\odot$) and
    quark core radius ($R_\text{Q,max}$, in km) at the maximum NS mass;
    radius of the $1.4 \text{M}_\odot$ ($R_{1.4 \text{M}_\odot}$, in km);
    mass of stars with central pressure equal to $P_\text{trans}$ ($M_\text{trans}$, in M$_\odot$).
    Results of the hybrid sets.
    }}
    \label{tab:NS_prop}
  \end{table*}

´

  \subsection{Tidal Deformability Result}

\begin{figure}[ht]
    \centering
    \includegraphics[width=\linewidth]{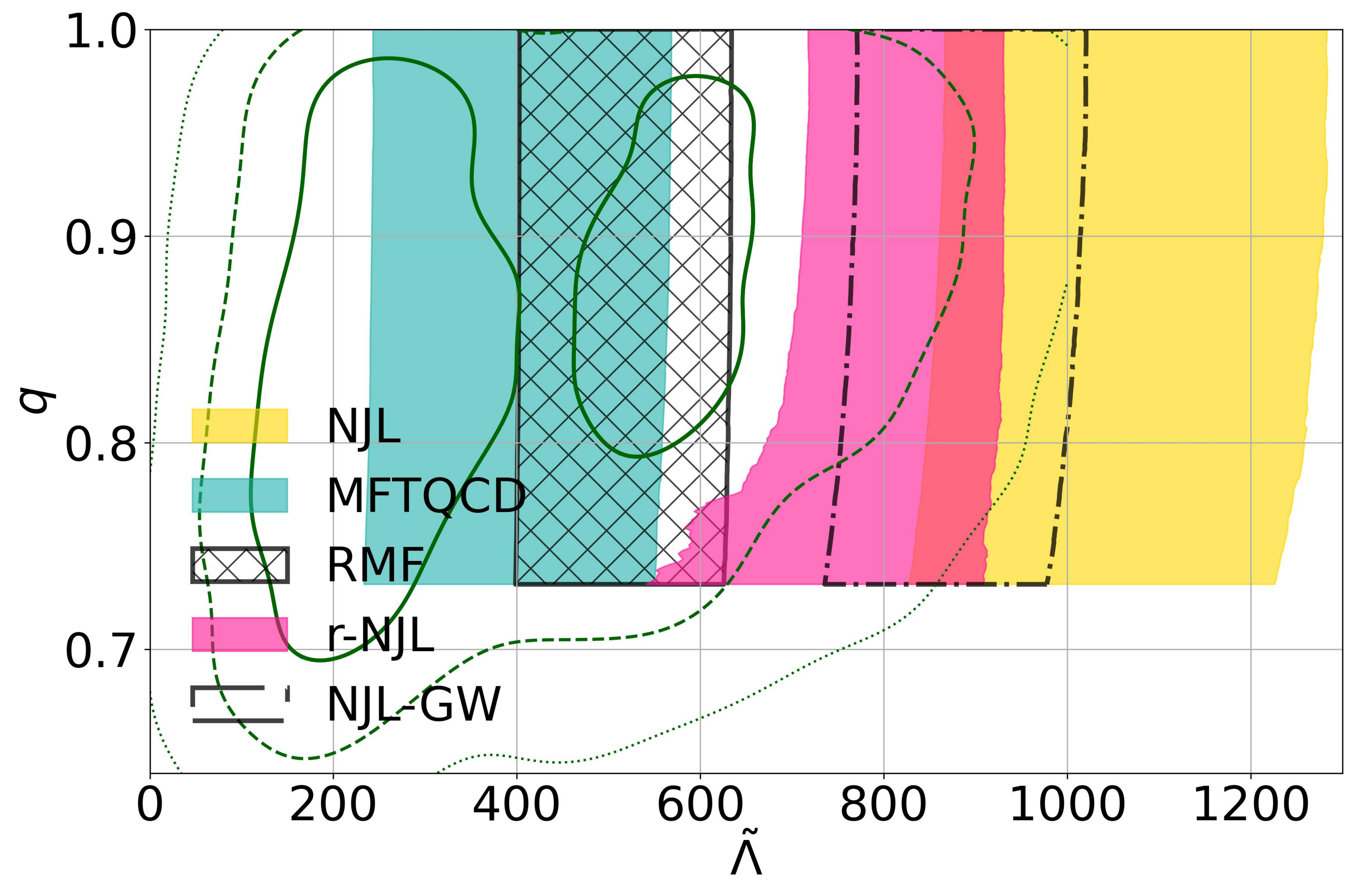}
    \caption{90\% of CI of tidal deformability.
      Same color code from Fig. \ref{fig:paper2_PxE}.
      The GW170817 observational data from the LIGO/Virgo collaboration is shown by
      solid (50\% CI), dashed (90\% CI) and dotted (99\% CI) green contours \cite{Abbott_2018}.}
    \label{fig:paper2_tidal}
  \end{figure}

  Fig. \ref{fig:paper2_tidal} shows the relation between the binary mass ratio
  $q = M_2/M_1 < 1$ and the effective tidal deformability, given by
  \begin{equation}
      \tilde{\Lambda} =
        \frac{16}{13} \frac{(12 q +1) \Lambda_1 +(12 +q) q^4 \Lambda_2}{(1 +q)^5)},
  \end{equation}
  where $\Lambda_i$ and $M_i$ are the tidal deformability and mass
  of the i-th NS in the binary system, respectively.
  For the different sets,
  $\tilde{\Lambda}$ values were calculated for
  $0.73 < q < 1$ and a fixed value of
  $M_\text{chirp} = (M_1 M_2)^{3/5} / (M_1 +M_2)^{1/5} = 1.186 \text{M}_\odot$,
  in accordance with event GW170817 \cite{Abbott_2018}.
  The observational data from the LIGO/Virgo collaboration
  for event GW170817 is shown in the figure by
  the solid green contours (50\% CI), dashed contours (90\% CI) and dotted contours  (99\% CI) \cite{Abbott_2018}.
  The NJL set does not satisfy the GW170817 constraint due to the large radii it predicts.
  Even though the GW170817 constraint was included in the Bayesian analysis,
  the NJL-GW set still struggles to describe the observational data.
  Nevertheless, the results for NJL-GW are within the 99\% CI of the data and
  are more favorable than those of the NJL set.
  The NJL equation has difficulty reconciling
  the tidal deformability data, which requires a softer equation,
  with the two solar mass data, which requires a stiffer equation.
  It is interesting to note that restricting the prior of the hadron parameters (r-NJL),
  has a greater effect on the $\tilde{\Lambda}$ values than imposing the GW170817 restriction (NJL-GW).
  Although the results of the r-NJL set only describe the GW170817 event at 99\% CI,
  it is the set that uses the NJL equation that best satisfies this observational data.
  However, it is also the set with the lowest maximum mass values,
  as we can see from Tables \ref{tab:NS_prop} and \ref{tab:NS_prop_RMF}.

  \subsection{Trace Anomaly Result}

  \begin{figure}[ht]
    \centering
    \includegraphics[width=.9\linewidth]{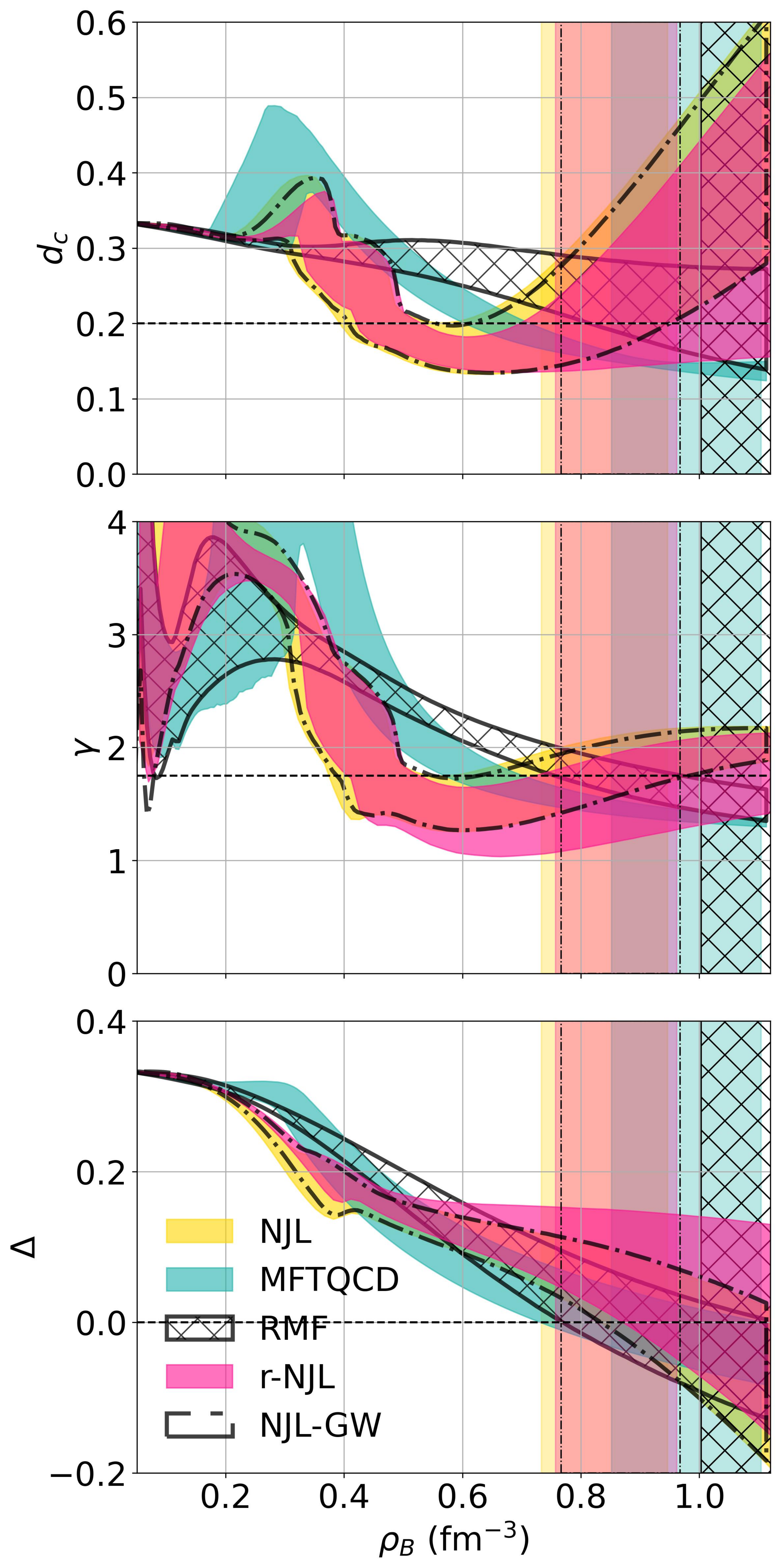}
    \caption{90\% of CI of measure of conformability ($d_c$),
      polytropic index ($\gamma$) and trace anomaly ($\Delta$)
      versus baryonic density.
      Same color code from Fig. \ref{fig:paper2_PxE}.}
    \label{fig:paper2_trace}
  \end{figure}

  Fig. \ref{fig:paper2_trace} shows
  the polytropic index $\gamma = d \ln P / d \ln \epsilon$,
  the trace anomaly $\Delta = 1/3 - P/\epsilon$,
  and the measure of conformability $d_c = \sqrt{\Delta^2 + \Delta'^2}$,
  with $\Delta' = d \Delta / d \ln \epsilon$, as functions of the baryonic density.
  These quantities were used in model-independent EOSs
  to  identify quark matter inside NSs
  \cite{Annala_2023}, \cite{Annala_2020}, \cite{Fujimoto_2022}.
  In \cite{Annala_2020}, it was proposed that $\gamma < 1.75$
  could indicate deconfined phase behavior.
  In \cite{Annala_2023}, $d_c < 0.2$ was used instead.
  However, none of our sets follow these trends.
  In $d_c$ plots, we can identify the phase transition
  by the bump at $\rho \approx 0.2 - 0.4 \text{ fm}^{-3}$ for the hybrid sets.
  After the phase transition, the value of $d_c$ decreases and
  only reaches $0.2$ at densities of approximately
  $\rho \approx 0.5 \text{ fm}^{-3}$ and $\rho \approx 0.7 \text{ fm}^{-3}$ 
  for the three NJLs and MFTQCD sets, respectively.
  Similar behavior is observed in the $\gamma$ plots.
  Furthermore, the hadron set reaches the $d_c$ and $\gamma$ limits at
  $\rho \approx 0.8 \text{ fm}^{-3}$,
  even though it contains no quarks.
  This may suggest that $d_c$ and $\gamma$ are not suitable quantities
  for indicating the presence of quark matter by themselves.
  All sets reach negative values of $\Delta$ at large densities.

  \begin{figure*}[ht]
    \centering
    \includegraphics[width=\linewidth]{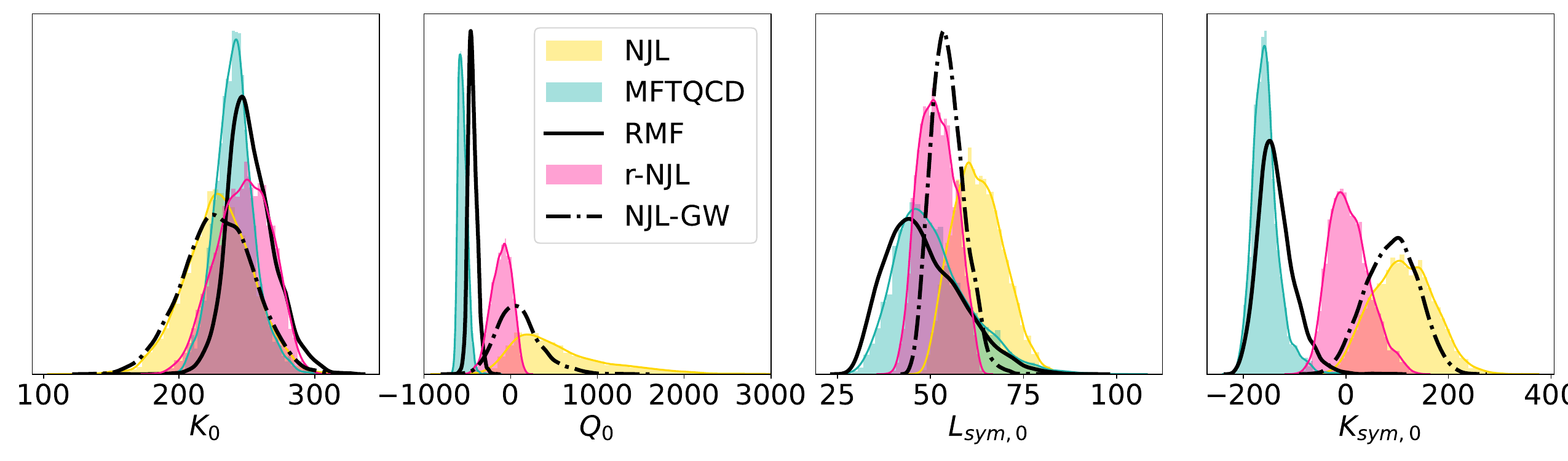}
    \caption{Histogram of nuclear matter properties for the NJL, MFTQCD, RMF, NJL-GW and r-NJL sets.}
    \label{fig:hist_nmp}
  \end{figure*}

\section{Discussion} \label{sec:Implications}

  In this section, we first discuss the NMP properties of the hadronic phase of the different sets. Several NS properties will then be compared in the following subsections, including the transition density to quark matter, the size of the quark core, the information that the slope of the mass-radius curve provides about the deconfinement phase transition, and the compactness of the maximum-mass configuration.

\subsection{Discussion of the nuclear matter properties}

  Fig. \ref{fig:hist_nmp} shows the  probability distributions  for some NMP obtained with the NJL, NJL-GW, r-NJL, MFTQCD and RMF sets.
 Tables \ref{tab:NMP_prop_RMF} and \ref{tab:NMP_prop} give the 90\% CI of NMP for the hybrid and hadronic sets, respectively.

The properties $\rho_0$, $EA$, $K_0$ and $J_{sym,0}$ have entered our Bayesian
inference as constraints (see  Table \ref{tab:nmp_constraints}).  Properties $\rho_0$, $EA$, and $J_{sym,0}$ 
show  distributions similar to the constraints imposed.  In  Fig. \ref{fig:hist_nmp}, we plot $K_0$, the only property included in the inference that deviates from the constraint imposed,  together with the properties not included in the inference $Q_0$, $L_{sym,0}$ and  $K_{sym,0}$. Concerning the
$K_0$ property: the histograms for models NJL and NJL-GW follow the
imposed constraint while all the others  show a shift of the peak to
larger values of $K_0$. This is due to the fact that   NJL and NJL-GW
are able to satisfy the two solar mass constraint imposed by the PSR
J0740 through the $\xi_{\omega\omega}$ term of the quark model while
the other scenarios rely completely or  strongly on the
hadronic EOS (RMF, MFTQCD and r-NJL).
\\

  \begin{table}[hb]
    \centering
    \setlength{\tabcolsep}{12.pt}
    \renewcommand{\arraystretch}{1.2}
    \begin{tabular}{cccc}
      \hline \hline
      \multicolumn{4}{c}{Set RMF} \\
      quant & median & min & max \\
      \hline
      $\rho_0$         & 0.159    & 0.158    & 0.161    \\
      EA               & -16.000  & -16.032  & -15.968  \\
      $K_0$            & 250.940  & 228.158  & 283.916  \\
      $Q_0$            & -445.834 & -510.475 & -346.521 \\
      $J_\text{sym,0}$ & 32.138   & 29.678   & 34.717   \\
      $L_\text{sym,0}$ & 46.898   & 34.300   & 66.925   \\
      $K_\text{sym,0}$ & -140.665 & -184.537 & -69.975  \\
      $Q_\text{sym,0}$ & 1136.710 & 407.590  & 1563.186 \\
      \hline
    \end{tabular}
    \caption{90\% CI of NMP of the hadronic RMF set.}
    \label{tab:NMP_prop_RMF}
  \end{table}
  
  \begin{table*}[ht]
    \centering
    \setlength{\tabcolsep}{12.pt}
    \renewcommand{\arraystretch}{1.2}
    \begin{tabular}{ccccccc}
      \hline \hline
      & \multicolumn{3}{c}{Set NJL} & \multicolumn{3}{c}{Set NJL-GW} \\
      quant & median & min & max & median & min & max \\
      \hline
      $\rho_0$         & 0.159   & 0.158    & 0.161    & 0.159    & 0.158    & 0.161    \\
      EA               & -16.000 & -16.033  & -15.967  & -16.000  & -16.032  & -15.968  \\
      $K_0$            & 229.623 & 188.253  & 269.753  & 228.829  & 185.733  & 271.410  \\
      $Q_0$            & 417.230 & -85.649  & 1663.222 & 71.412   & -273.811 & 573.546  \\
      $J_\text{sym,0}$ & 31.829  & 29.363   & 34.470   & 31.149   & 28.706   & 33.708   \\
      $L_\text{sym,0}$ & 62.443  & 52.668   & 74.218   & 54.256   & 48.117   & 62.705   \\
      $K_\text{sym,0}$ & 110.178 & 17.644   & 206.472  & 91.128   & 6.448    & 169.655  \\
      $Q_\text{sym,0}$ & 902.673 & -426.280 & 1476.560 & 1002.040 & 202.071  & 1506.580 \\
      \hline \hline
      & \multicolumn{3}{c}{Set MFTQCD} & \multicolumn{3}{c}{Set r-NJL} \\
      quant & median & min & max & median & min & max \\
      \hline
      $\rho_0$         & 0.159    & 0.158    & 0.161    & 0.159    & 0.158    & 0.160    \\
      EA               & -16.000  & -16.034  & -15.967  & -16.000  & -16.032  & -15.968  \\
      $K_0$            & 240.731  & 218.267  & 266.119  & 248.612  & 213.777  & 278.984  \\
      $Q_0$            & -557.802 & -619.236 & -462.206 & -97.085  & -307.695 & 73.622   \\
      $J_\text{sym,0}$ & 32.563   & 30.206   & 35.048   & 31.473   & 28.863   & 33.655   \\
      $L_\text{sym,0}$ & 49.115   & 36.964   & 69.491   & 51.564   & 44.450   & 59.747   \\
      $K_\text{sym,0}$ & -158.530 & -191.779 & -106.057 & 2.801    & -49.118  & 78.403   \\
      $Q_\text{sym,0}$ & 976.226  & 337.799  & 1475.125 & 1276.510 & 906.791  & 1557.227 \\
      \hline
    \end{tabular}
    \caption{90\% CI of NMP of the hybrid sets: NJL, NJL-GW, MFTQCD and r-NJL.}
    \label{tab:NMP_prop}
  \end{table*}


Concerning the other properties (the isoscalar $Q_0$ and the isovector
$L_{sym,0}$, $K_{sym,0}$), the following comments are
in order: i)  MFTQCD and RMF show similar distributions, with a
tendency of MFTQCD to have slightly smaller properties. This can be
understood because the hadron
phase properties of MFTQCD are defined for densities of the order
of 0.15 - 0.25 fm$^{-3}$, which are close to the density that defines the
NMP and the densities at which the  PNM were imposed in the Bayesian
inference.  MFTQCD shows a slightly softer behavior because the
stiffness required to describe a two solar mass system is defined by the quark phase;  ii) For  $Q_0$ and
$L_{sym,0}$, $K_{sym,0}$,  the NJL-GW set and (more strongly) the r-NJL set show NMP
distributions intermediate between the more extreme values of NJL  and
the ones of RMF and MFTQCD. Including the GW constraint and
restricting the parameter prior to the one used for the RMF model
softens the EOS, shifting these three parameters to smaller values
compared to NJL; iii) The MFTQCD and RMF sets prefer negative values
for $K_{sym,0}$,  in agreement with \cite{Li_2021}. However, the NJL
and NJL-GW sets take positive values of $K_{sym,0}$. A positive and
large value was also obtained in other analyses, see for instance
\cite{PhysRevC.109.035803}, where a RMF model that also includes the $\delta$-meson has been fitted
to the CREX and PREX data; iv) The symmetry energy slope $L_{sym,0}$
in all models is consistent with $L_{sym,0}  = 58.7\pm 28.1$~MeV
obtained at 1$\sigma$ from both experimental and observational data
\cite{Oertel_2017}
\\
In summary: the NMP constrained experimentally ($EA$, $\rho_0$,
$K_0$, $J_{sym,0}$  and $L_{sym,0}$)  take values within the expected
interval in all scenarios.
The properties obtained for the hadron matter of the hybrid stars
reflect the condition that was imposed to generate large quark cores and still describe two solar masses:  the deconfiment occurs
between $\sim\rho_0$ and $3 \rho_0$. 
  The chiral symmetric model NJL with the vector contribution gives a stiff quark EOS. An early
  phase transition is only possible if the hadron phase is also   stiff
  at high densities and, therefore, the higher order terms in the
  expansion of the energy per particle, eq. \ref{eq:X_nmp} and \ref{eq:X_nmp_sym}, take large values.
  MFTQCD is a  model that describes matter already in a chiral
  symmetric phase. By controlling the vector interaction and the bag
  contribution it is easy to identify a region in parameter space   that
  satisfies the conditions imposed    without stiffening the hadron
  EOS.
\\

  \subsection{Deconfinment phase transition}

  \begin{figure}[ht]
    \centering
    \includegraphics[width=.8\linewidth]{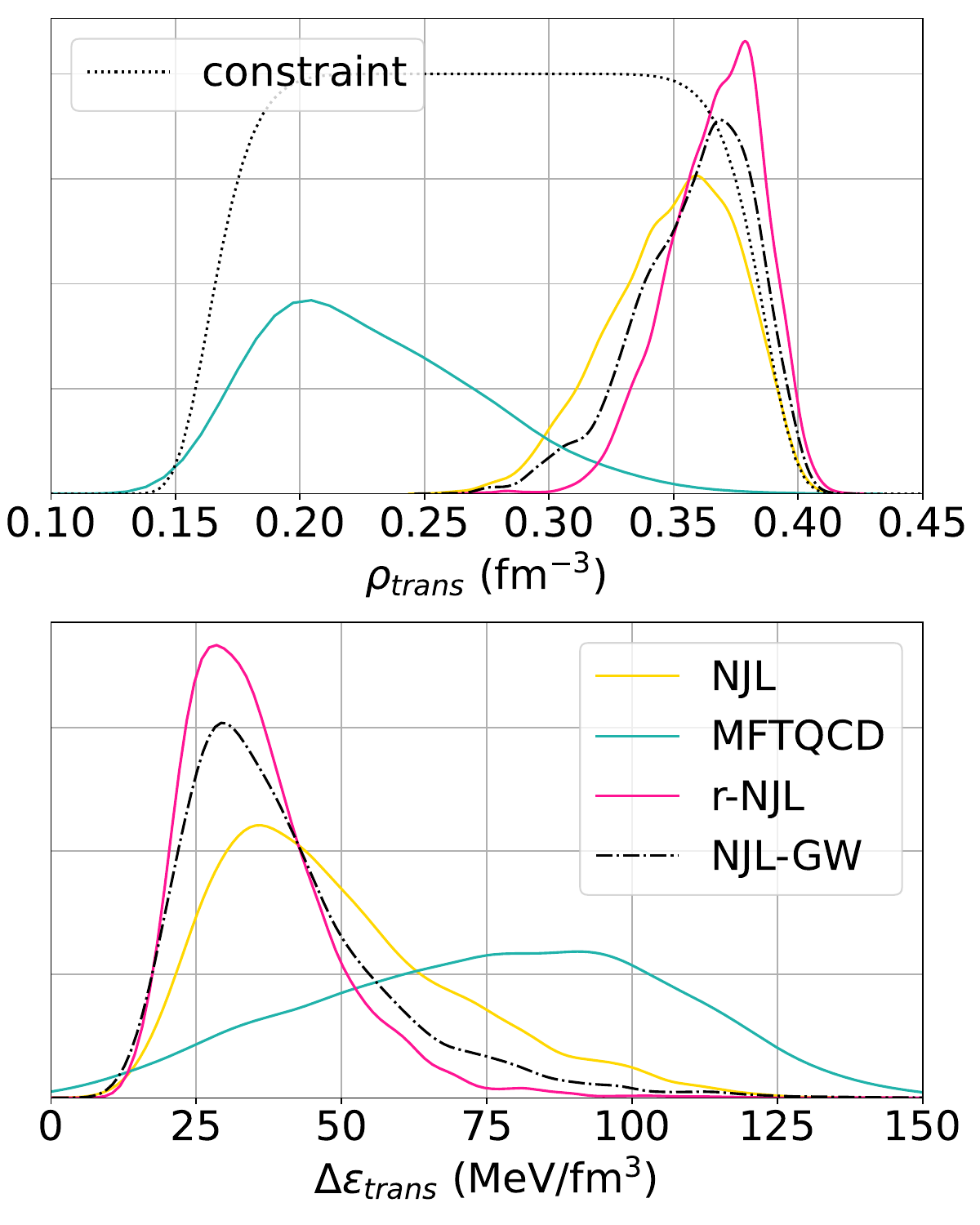}
    \caption{Histogram of $\rho_\text{trans}$ and $\Delta \epsilon_\text{trans}$
      for the NJL, MFTQCD and NJL-GW sets.}
    \label{fig:hist_phT_prop}
  \end{figure}

    \begin{figure}[ht]
    \centering
    \includegraphics[width=.8\linewidth]{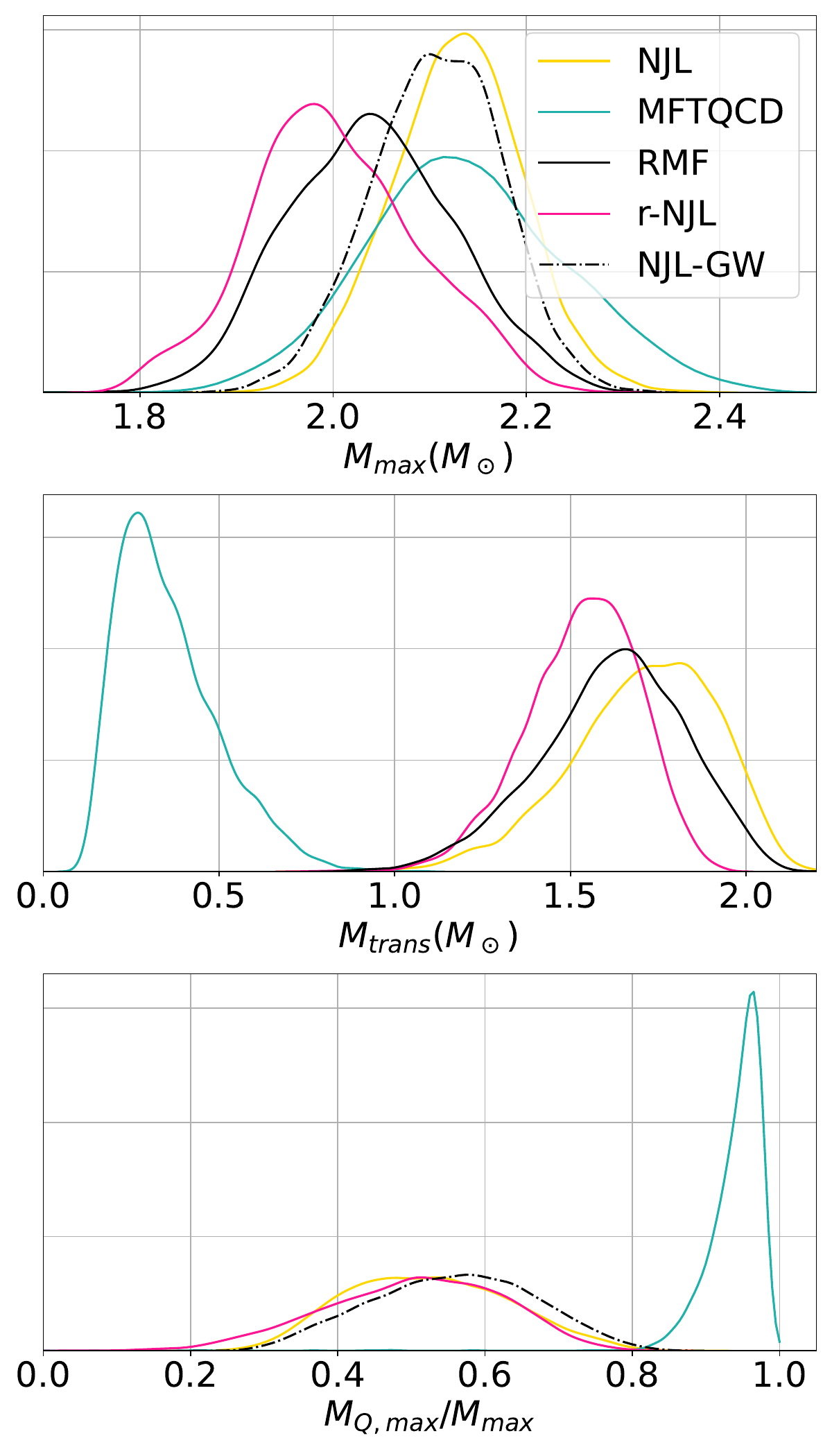}
    \caption{Histogram of $\rho_\text{trans}$ and $M_\text{max}$
      for the NJL, MFTQCD, RMF, NJL-GW and r-NJL sets.
      The maximum M$_\text{max}$ reached by each set is
      $2.384 M_\odot$,
      $2.444 M_\odot$,
      $2.302 M_\odot$,
      $2.327 M_\odot$,
      $2.291 M_\odot$,
      respectively.
      }
    \label{fig:hist_M_prop}
  \end{figure}

    \begin{table}[hb]
    \centering
    \setlength{\tabcolsep}{7.pt}
    \renewcommand{\arraystretch}{1.2}
    \begin{tabular}{cccc}
      \hline \hline
                                                     & NJL-GW & NJL  & MFTQCD \\
      \hline
      $M_\text{trans} < 0.5 \; \text{M}_\odot$       &    0   &    0 & 4113   \\
      $0.5 < M_\text{trans} < 1.4 \; \text{M}_\odot$ & 1112   &  536 &  766   \\
      $M_\text{trans} > 1.4 \;  \text{M}_\odot$      & 6409   & 5978 &    0   \\
      total                                          & 7521   & 6514 & 4879   \\
      \hline
    \end{tabular}
    \caption{For each EOS set, the number of equations that fall in the bands  $M_\text{trans} < 0.5 \; \text{M}_\odot$, $0.5 < M_\text{trans} < 1.4 \; \text{M}_\odot$  and $M_\text{trans} > 1.4 \;  \text{M}_\odot$. In Fig.  \ref{fig:Mtrans} the speed of sound behavior and mass-radius curves for the two bands that are populated are shown for the NJL and MFTQCD sets.}
    \label{tab:n_eq_Mtrans}
  \end{table}
  
  \begin{figure}[ht]
    \centering
    \includegraphics[width=0.9\linewidth]{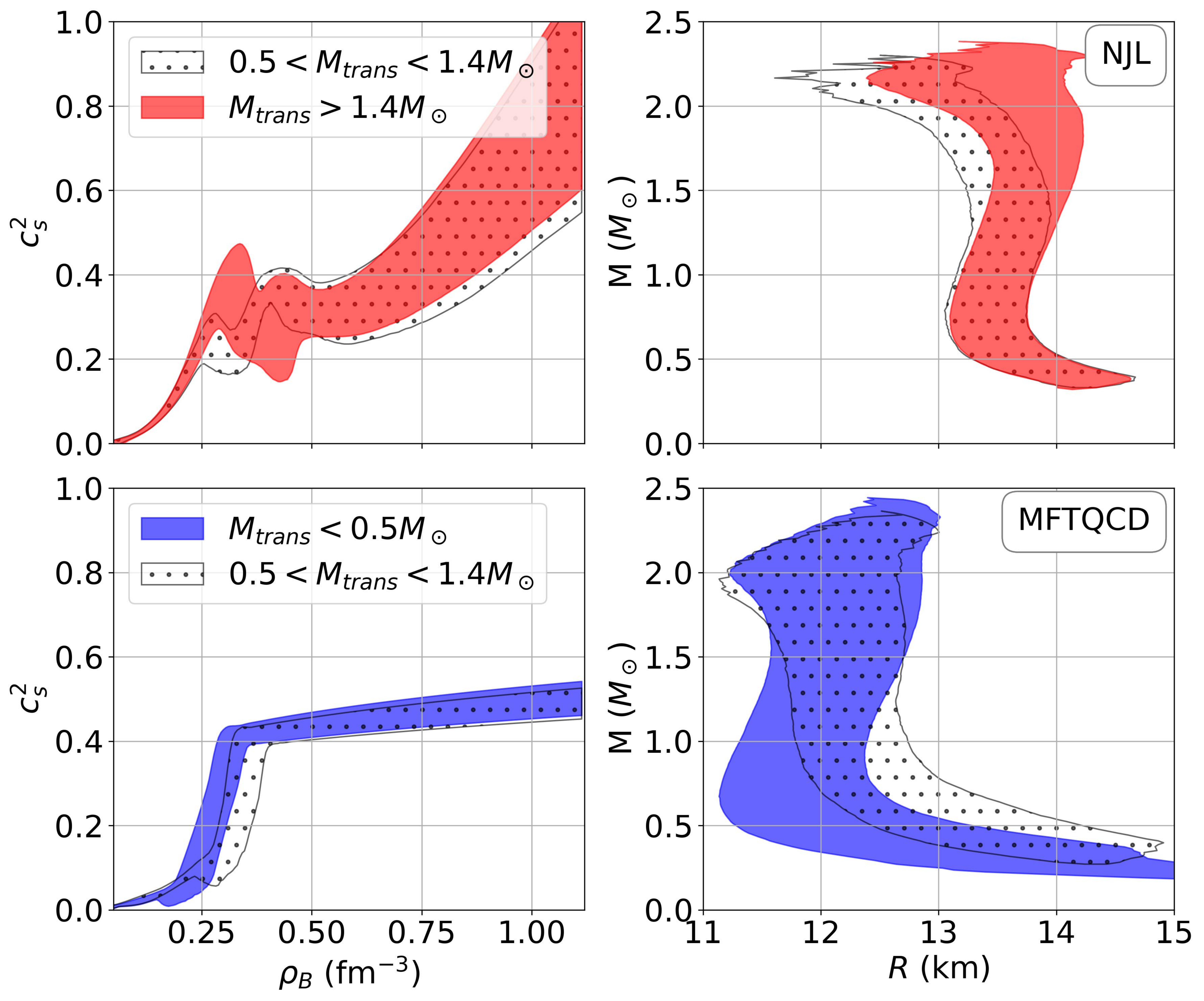}
    \caption{Speed of sound as a function of density (left) and
      mass-radius diagram (right) 
      for the NJL (top) and MFTQCD (bottom) sets.
      90\% CI of equations with $M_\text{trans} < 0.5 \text{M}_\odot$ (blue, only MFTQCD)
      $0.5 \text{M}_\odot < M_\text{trans} < 1.4 \text{M}_\odot$  (white dotted, MFTQCD and NJL) and
      $M_\text{trans} > 1.4 \text{M}_\odot$ (red, only NJL)  are represented.}
    \label{fig:Mtrans}
  \end{figure}

  In this subsection the transition density to a quark phase and other properties of the hybrid stars obtained within the two quark models are compared and discussed.\\
Probability distributions of the $\rho_\text{trans}$ (top plot) and   the strength of the phase transition defined as  
  $\Delta \epsilon_\text{trans} = \epsilon_\text{Q, trans} -\epsilon_\text{H,trans}$  (lower plot)
  are shown in Fig. \ref{fig:hist_phT_prop}.
 The sets using the NJL model allow phase transitions only above
  $\sim 0.25 \text{ fm}^{-3}$ (all NJL sets)
  while the MFTQCD set allows a phase transition within the entire allowed range.
 However, for the constraints imposed, the MFTQCD set prefers a lower phase transition
  with a median of $\rho_\text{trans} = 0.222 \text{ fm}^{-3}$,
  while the NJL sets prefer a higher phase transition density, with a median of
  $\rho_\text{trans} \sim  0.35-0.37 \text{ fm}^{-3}$.
 The behavior of the NJL models with respect to that of the MFTQCD results from the fact that the NJL EOSs are stiffer, favoring a late phase transition.\\
 The predictions of the two quark models can also be compared through  the strength of the phase transition 
  $\Delta \epsilon_\text{trans} $.
 The $\Delta \epsilon_\text{trans}$ distribution
  has the widest distribution for the MFTQCD set
  and  peaks  at larger $\Delta \epsilon_\text{trans}$, indicating that the strongest phase transitions occur with  the MFTQCD.
 The  NJL, NJL-GW and r-NJL sets show much weaker phase transitions  about three times weaker than the corresponding $\Delta \epsilon_\text{trans}$ for MFTQCD.\\
 Fig. \ref{fig:hist_M_prop} shows the histogram of
  the maximum NS mass (top),
  {the mass of the star for which quarks start to nucleate, i.e.} with $P_c = P_\text{trans}$ (middle) and the quark core mass of the maximum mass NS (lower).
 All hybrid stars reach  similar  maximum mass values
  with a median of $M_\text{max} \sim 2.13 \text{M}_\odot$, while the RMF set  median is 0.1$M_\odot$ smaller.
 The largest masses occur with MFTQCD at approximately 2.4 $M_\odot$, which is about 0.1 $M_\odot$ larger than all other sets. The largest masses are obtained for hybrid stars because the properties of the hadronic EOS (RMF) are strongly constrained by the NMP. In contrast, the quark EOS is only constrained by causality and pQCD, allowing for more freedom. This could suggest that very massive stars are a clear indication of deconfined quark matter within their interiors. \\
 A property that clearly distinguishes the two quark models is the hybrid star mass at the deconfinement transtion. 
  Within the  MFTQCD set, quark matter can exist in NSs with very low masses, well below 1~$\text{M}_\odot$, the median value  being equal to 0.327~$\text{M}_\odot$, see middle panel of  Fig. \ref{fig:hist_M_prop}. These are essentially quark stars with a hadronic crust,
  allowing NSs with a quark core mass larger than $2 \text{M}_\odot$.
 NJL models predict the presence of quarks in heavier stars with a mass above 1$M_\odot$, with the median lying above $\sim$1.5$M_\odot$.   For NJL sets
 $M_\text{Q,max}^\text{med} \sim 1.1 \text{M}_\odot$, so the maximum quark core is about half of the maximum NS mass.
    To distinguish between these scenarios, one would expect other NS properties, such as those obtained from cooling, NS modes, or binary neutron star mergers, to reveal the different compositions. \\

  \begin{table*}[ht]
    \centering
    \setlength{\tabcolsep}{12.pt}
    \renewcommand{\arraystretch}{1.2}
    \begin{tabular}{ccccccccc}
      \hline \hline
      $M$ & \multicolumn{2}{c}{$1.2 \text{M}_\odot$} & \multicolumn{2}{c}{$1.4 \text{M}_\odot$}
        & \multicolumn{2}{c}{$1.6 \text{M}_\odot$} & \multicolumn{2}{c}{$1.8 \text{M}_\odot$} \\
      $dM/dR$ & + & - & + & - & + & - & + & - \\
      \hline
      NJL    & 6306 &  208 & 6008 &  506 & 4896 & 1618 & 2509 & 4005 \\
      NJL-GW & 7268 &  253 & 6656 &  865 & 4699 & 2822 & 1653 & 5868 \\
      r-NJL  & 5180 &  147 & 4562 &  765 & 2358 & 2969 &  206 & 5121 \\
      MFTQCD & 4392 &  487 & 4326 &  553 & 3459 & 1420 & 1119 & 3760 \\
      RMF    &  822 & 5215 &  175 & 5862 &   70 & 5967 &   38 & 5999 \\
      \hline
    \end{tabular}
    \caption{Number of EOSs with positive/negative slope in the mass-radius diagram at
        $M = 1.2, 1.4, 1.6, 1.8 \text{M}_\odot$.}
    \label{tab:slope_MR}
  \end{table*}
  
To better understand the effect of the deconfinement phase transition on the NS properties,  we have divided each EOS set into two bands according to the value of the  $M_\text{trans}$. The EOS of set MFTQCD populate two bands $M_\text{trans} < 0.5 \text{M}_\odot$ and  $0.5 \text{M}_\odot < M_\text{trans} < 1.4 \text{M}_\odot$ while NJL models populate bands  $0.5 \text{M}_\odot < M_\text{trans} < 1.4 \text{M}_\odot$ and  $M_\text{trans} > 1.4 \text{M}_\odot$.
Fig. \ref{fig:Mtrans} shows the behavior of the  speed of sound (left) and  the mass-radius curves (right) for the  NJL (top) and the MFTQCD (bottom) quark models within each band.  Table \ref{tab:n_eq_Mtrans} gives the number of equations in each band: most of the MFTQCD (NJL and NJL-GW) stars have $M_\text{trans}<0.5\, \text{M}_\odot$ ($M_\text{trans}>1.4\, \text{M}_\odot$).\\
The first peak in the speed of sound occurring for the NJL set (other NJL sets have a similar behavior) is due to the deconfinement phase transition which occurs at lower densities for the band   $M_\text{trans}<0.5\, \text{M}_\odot$.  The second  peak identifies the onset of the $s$-quark and occurs at the same density in both bands because it is not affected by the deconfinement transition. For the MFTQCD a phase transition at low densities imposes a faster rise of the speed of sound at low densities.
  In the mass-radius diagram, smaller  $M_\text{trans}$ values correspond to smaller radii for low mass stars in the MFTQCD set and large mass stars in the NJL set.
 Some conclusions are in order: i)  only the  MFTQCD set with a deconfinement phase transition at low densities,   $M_\text{trans} < 0.5 \text{M}_\odot$, is able to describe the HESS data, as found in other works \cite{DiClemente:2022wqp,Brodie:2023pjw,Sagun:2023rzp,Laskos-Patkos:2023tlr,Ayriyan:2025rub,Blomqvist:2025cxe}; ii) both the sets built with MFTQCD and with NJL models  suggest that PSR J0740+6620 could be a hybrid star; iii) no conclusive statement is drawn for pulsar PSR J0030+0451: it has a large (small or none)  quark inner core within the MFTQCD (NJL) model. Other observations as binary NS mergers may be able to identify a large quark core \cite{Bauswein2018}.

  \subsection{Mass-radius curve slope}

    We next analyze the slope of the  mass-radius curve. In \cite{Ferreira:2024hxc}, it has been shown that this quantity reflects the behavior of some NS properties such as their maximum mass and radius. In addition, it  has been discussed in \cite{Ferreira:2025dat} that the composition could affect the sign of the slope.
    Table \ref{tab:slope_MR} shows the number of EOSs with positive and negative slope, i.e. $dM/dR$, at
    $M = 1.2,\, 1.4, \, 1.6,\, 1.8 \text{M}_\odot$.
    The RMF set is the set that presents the largest number of negative slope values for all masses, above $\sim97\%$ for all masses except for 1.2$M_\odot$ which is still of the order of 86\%. 
    The hybrid sets mostly have positive values for $M = 1.2, 1.4, 1.6$ $\text{M}_\odot$, around or above 90\%. 
    For $M = 1.8\, \text{M}_\odot$,  there are still about 20\%-38\% EOSs with a positive slope. Only r-NJL shows a smaller value, $\sim 3\%$ in line with the RMF set. \\
    Similar conclusions were drawn in reference \cite{Ferreira:2025dat} for the nucleonic EOS: essentially, a negative slope was obtained for all masses. 
    For the hyperonic EOSs, positive slope values were mostly obtained for small masses,  similarly to the results obtained for the hybrid sets in this study. In both studies,
    a positive value of $dM/dR$ at small masses could indicate the presence of exotic degrees of freedom, such as quarks or hyperons. \\
    A significant difference in our present study is that, even at $1.8  \text{M}_\odot$, a reasonable fraction of the curves still have a positive slope. This reflects the two models used to describe hybrid stars, with the quark EOSs being less constrained. The vector terms are responsible for the stiff EOSs that justify the positive slope for this large mass. If it is confirmed in the future that two solar mass stars, such as  J0740--6620 (with $R=12.48^{+1.28}_{-0.88}$  at 68\% CI \cite{Salmi_2024})  have a larger radius  than 1.4 solar mass stars such as the radius of the pulsar PSR J0614--3329 (with $R=10.29^{+1.01}_{-0.86}$  at 68\% CI \cite{Mauviard:2025dmd}),  then a quark core of strongly interacting matter could explain the difference in radii. A positive slope for masses of about 1.8 solar masses was also obtained in the agnostic description of the EOS undertaken in in \cite{Ecker:2024uqv}.

  \subsection{Maximum compactness}
  
  \begin{figure}[ht]
    \centering
    \includegraphics[width=\linewidth]{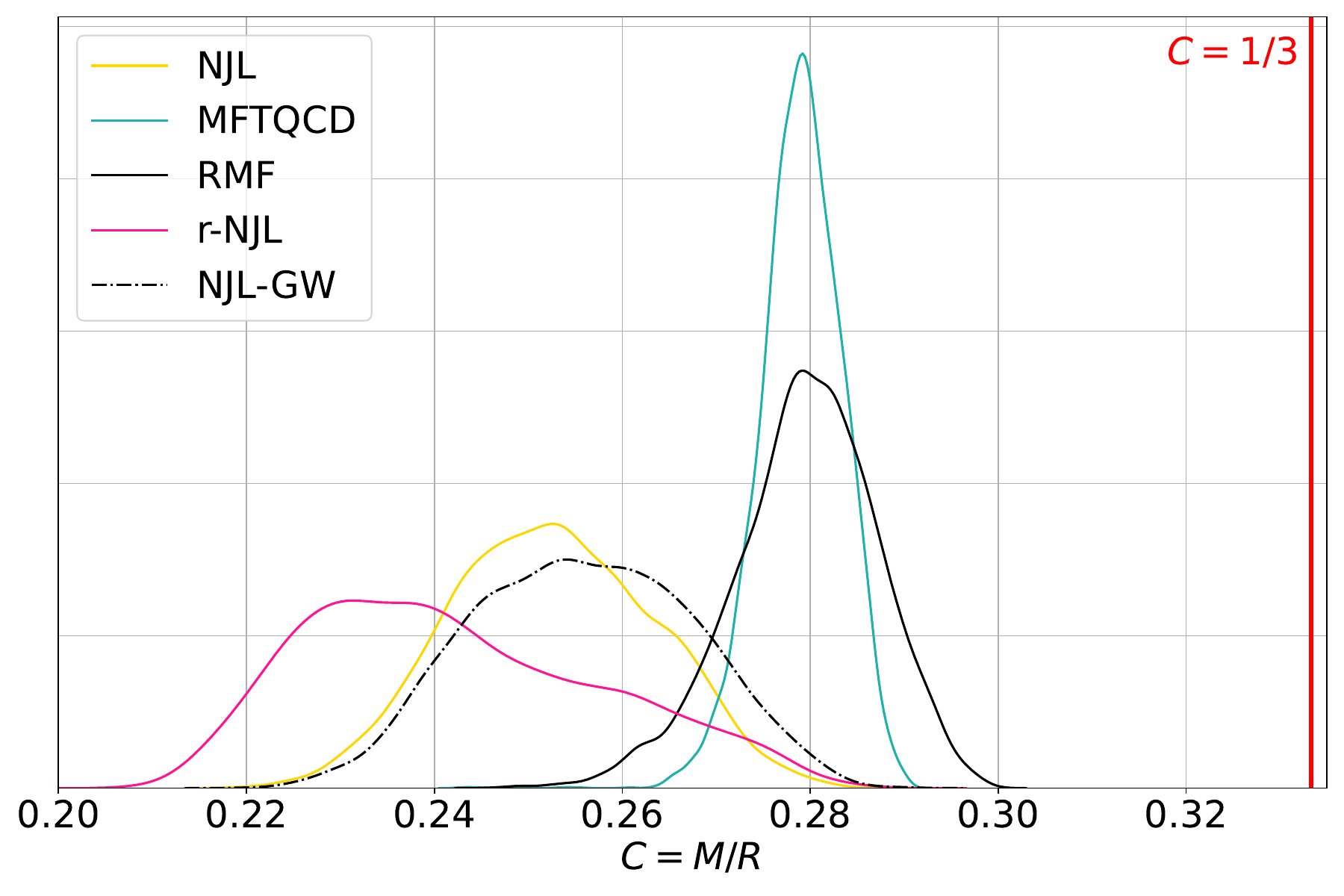}
    \caption{Compactness $\mathcal{C} = M_\text{TOV}/R_\text{TOV}$ distribution for all sets:
        NJL-GW (purple), NJL (orange), MFTQCD (cyan), r-NJL (blue) and RMF (black dashed).
        Upper limit $\mathcal{C}_\text{max}= 1/3$ \cite{Rezzolla:2025pft} is shown as the vertical red line.}
    \label{fig:C}
  \end{figure}

  In \cite{Rezzolla:2025pft}, the authors discuss the maximum compactness of NSs. Considering an agnostic description of the EOS and imposing both observational and theoretical constraints, they  conclude that if the maximum compactness $\mathcal{C}_{max}$ is attained by the star with the maximum mass of the sequence of nonrotating configurations, i.e. $\mathcal{C}_\text{TOV}= M_\text{TOV}/R_\text{TOV}$,  then $\mathcal{C}_\text{max}\le 1/3$.  In our study, we consider EOS motivated by microscopic models and impose similar constraints \footnote{Note that we do not include the pulsar PSR J0952-0607 in our analysis  due to the large uncertainty associated with its mass.}. 
  Although the extremes are not attained because the models are not general enough, it is interesting to identify the maximum values obtained for  $\mathcal{C}$ in our five datasets (Fig. \ref{fig:C}) and analyze the properties of the corresponding stars.     Additionally, Table \ref{tab:C} shows the the minimum and maximum values of  $\mathcal{C}_\text{TOV}$ for each set.

    The NJL sets have the lowest compactness values because they have smaller maximum masses and larger radii.    The MFTQCD and RMF sets have the largest values of  $\mathcal{C}$, reaching 0.290 and 0.299, respectively  (see Table \ref{tab:C}). This is because the maximum-mass stars in these sets have much smaller radii and larger masses.
    All sets satisfy the $\mathcal{C} \leq 1/3$ constraint, represented by the vertical red line,  obtained in \cite{Rezzolla:2025pft}. In \cite{Lattimer:2019eez}, a maximum compactness of 0.354 was obtained by considering a soft EOS  at low densities and an EOS that is as stiff as possible, yet still causal, at high densities. In \cite{Rezzolla:2025pft}, pQCD constraints have resulted in a slightly smaller maximum compactness.

  \begin{table}[hb]
    \centering
    \setlength{\tabcolsep}{7.pt}
    \renewcommand{\arraystretch}{1.2}
    \begin{tabular}{ccc}
      \hline \hline
      Set & $C_\text{min}$ (M/R) & $C_\text{max}$ (M/R) \\
      \hline
      NJL-GW & 0.219 (1.93/13.00) & 0.291 (2.29/11.46) \\
      NJL    & 0.219 (1.96/13.17) & 0.288 (2.21/11.31) \\
      MFTQCD & 0.263 (1.86/10.41) & 0.290 (2.44/12.39) \\
      r-NJL  & 0.209 (1.87/13.18) & 0.286 (2.27/11.72) \\
      RMF    & 0.246 (1.87/11.17) & 0.299 (2.26/11.17) \\
      \hline
    \end{tabular}
    \caption{Minimum and maximum values of $\mathcal{C}_\text{TOV}$ for each set.
        In parentheses, we have the respective values of mass and radius in M$_\odot$ and km.}
    \label{tab:C}
  \end{table}

\begin{figure}[ht]
    \centering
   \includegraphics[width=\linewidth]{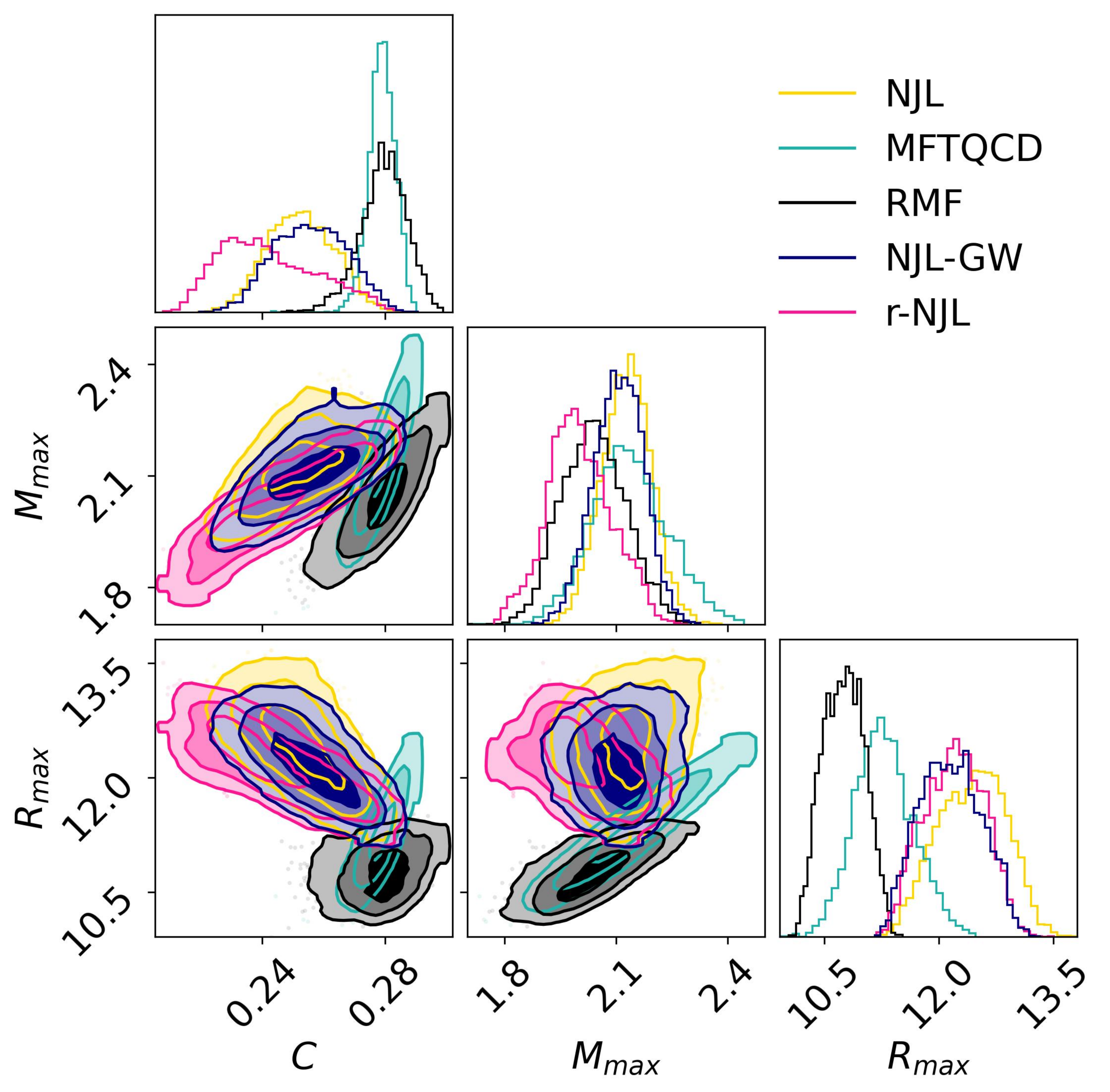}
    \caption{Corner plot of the compactness ($\mathcal{C}$),       maximum mass ($\Mmax$) and radius at the maximum mass ($R_\text{max}$).}
   \label{fig:cornerC}
  \end{figure}

\begin{figure}[ht]
    \centering
    \includegraphics[width=\linewidth]{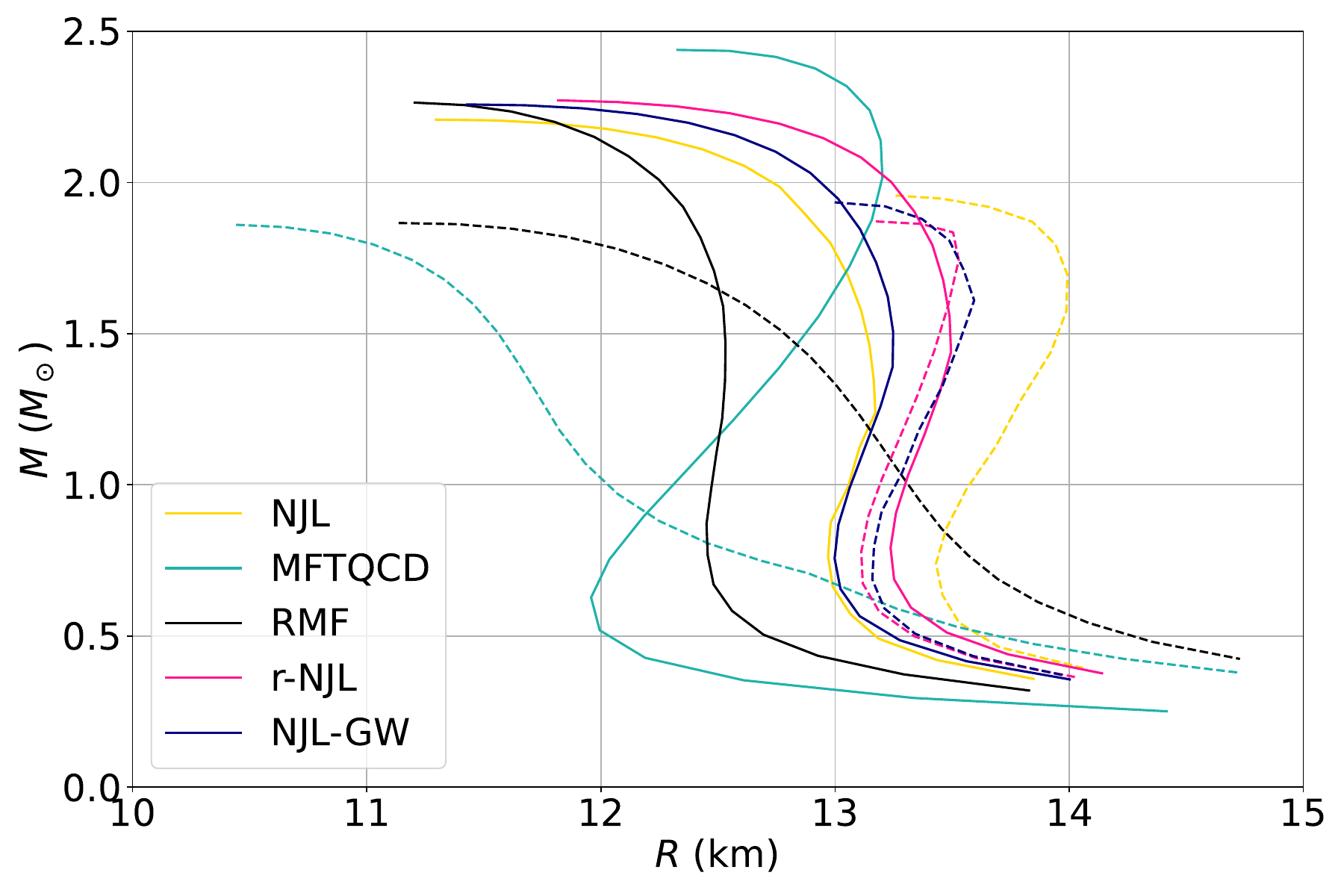}
    \caption{Mass-radius diagram of the maximum (solid lines) and minimum (dashed lines)
        of $\mathcal{C} = M_\text{max}/R_\text{max}$.}
    \label{fig:Cmax}
  \end{figure}

    To understand better the behavior of the $\mathcal{C}$, a corner plot with $\mathcal{C}$, $M_\text{max}$, and $R_\text{max}$ is given in Fig. \ref{fig:cornerC}.
    The $\mathcal{C} \times M_\text{max}$ panel shows that there is a correlation between these two quantities: larger values of $M_\text{max}$ result in larger values of $\mathcal{C}$ as expected. 
    The radius of the maximum mass configuration of NJL sets shows no correlation with the maximum mass, while for the RMF and MFTQCD the radius increases with the mass. We have plotted in Fig. \ref{fig:Cmax} the mass-radius curves corresponding to 
    the maximum (solid) and minimum (dashed) of $\mathcal{C}$, for each set. The maximum mass significantly impacts the definition of the maximum $\mathcal{C}$. 
  
\section{Conclusions} \label{sec:Conclusion}

    This work explores the possibility of quark matter existing within neutron stars (NSs).
    We derived the EOSs using microscopic models. As a result,  the composition of these objects is known, and it is possible to discuss the effect of the presence of a quark core.
    
    We used the relativistic mean field (RMF) model to describe the hadron phase of matter. This is a framework that has been quite successful in describing  the properties of finite nuclei and nuclear matter.    For the quark phase, we employed the Nambu–Jona-Lasinio (NJL) model and the mean-field theory of quantum chromodynamics (MFTQCD).
    MFTQCD is an MIT like model that describes quarks with current masses in a chiral symmetric phase and includes gluon contributions through a vector contribution and a bag pressure. Within this model, a  deconfinement phase transition at low densities is possible. The NJL model includes chiral symmetry breaking and restoration and the quark masses are dynamical.  In addition to the usual four quark vertices, we have  also  included  eight quark vertices in the vector sector, which allow for large NS masses \cite{Benic:2014iaa}. This model does not predict a deconfinement phase transition below twice saturation density.
    We applied a Maxwell construction to describe the quark-hadron phase transition.
    Using Bayesian inference with theoretical constraints from nuclear matter properties and pQCD calculations, as well as observational constraints from NICER and GW170817, we obtained large sets of EOSs. 

   In total, five sets of EOSs were obtained: four with hybrid EOSs and one with nucleonic EOSs (the RMF set).  Three of the hybrid sets used the NJL model for the quark phase. Two of these sets, NJL and NJL-GW, differed in that the latter was also constrained by the GW170817 observation. Additionally, a third NJL set was built to examine the impact of the nucleonic phase prior.  The MFTQCD model was considered in the fourth hybrid EOS set to describe the quark core.

     
    We have identified the regions in the pressure-energy density plane spanned by the different sets and  compared them to the expectations from agnostic approaches, as discussed in \cite{Annala:2021gom,Altiparmak:2022bke}. While the RMF EOS distribution follows the 90\% CI predicted with an agnostic approach \cite{Altiparmak:2022bke}, the hadron phase of the NJL sets lies above the   90\% CI band, while the quark phase is consistent with the 90\% CI of the agnostic study.  An early phase transition is only possible for a stiff quark EOS, such as the NJL EOS, if the hadron phase is also stiff. Finally, the MFTQCD EOS probabilities of low-mass hybrid stars populate the low-energy density region below the 90\% CI of the agnostic study. This seems to indicate that a low-density deconfinement phase transition is necessary to cover this region.

    In order to obtain large quark cores, we have imposed a constraint on the phase transition density value,  allowing it to occur at not too high densities, in particular,  in the range
    $0.16 \lesssim \rho_\text{trans} \lesssim 0.40 \text{ fm}^{-3}$.
    MFTQCD set can describe NSs with a quark core mass up to  $\sim 2 \text{M}_\odot$,
    while for the NJL sets the  phase transition occurs above $2 \rho_0$,
    and the maximum  quark core mass is just above $ 1 \text{M}_\odot$. The main difference between the two approaches is the quark masses at deconfinement. The dynamical quark masses in NJL description are still far from their current masses at deconfinement. Moreover, the onset of the s-quark occurs after deconfinement. On the contrary, within the MFTQCD descritpion at deconfinement all three quarks are present and are described by their current quark masses.
    One consequence of the different transition densities is that the MFTQCD set predicts the existence of quark matter in the core of a 1.4 $M_{\odot}$ NS, while the NJL  sets favor EOSs without quark matter inside these NS.  However, note that NJL sets do not completely rule out this possibility, as a transition is possible for stars with a mass above approximately $1.3 \text{M}_\odot$, though median values favor quark matter in stars with a mass above $1.7 \text{M}_\odot$. This reflects the restricted flexibility of the NJL framework. A quark model that combines the chiral dynamics of the NJL with greater flexibility at low densities would be valuable to explore in future work. 
    All hybrid sets predict that  quark matter is present  inside  NSs with masses greater than 2 M$_\odot$, such as the pulsar  pulsar J0740--6620.
    


    Concerning the maximum mass, the RMF set has the smallest $M_\text{max}$, below  $2.2\text{M}_\odot$ (90\% CI). 
     The presence of quarks  in sets  MFTQCD, NJL and NJL-GW  results in slightly larger maximum masses above $\approx 2.2 \text{M}_\odot$ (90\% CI).
The  absolute maximum mass has been attained by the MFTQCD model above 2.4$M_\odot$, while all other models have maximum masses below 2.3$M_\odot$.
    Hybrid EOS reach larger  maximum masses than the RMF EOS because there is some freedom in building the quark phase, which is only constrained by causality and the pQCD constraints, along with a maximum mass of at least two solar masses. Both quark models considered contain vector contributions that can stiffen the EOS inside NSs  while still satisfying the pQCD constraints at very large densities. This stiffening could explain the larger radius predicted for the two solar mass pulsar J0740--6620  compared with the  radius of the  1.4 solar mass pulsar J0614--3329.

The slope of the mass-radius curve may carry information about the possible presence of exotic matter inside NSs. If the slope of the mass curve is negative for all masses, then there is a high probability that the NS does not contain exotic matter. The opposite conclusion is drawn when the slope is positive at 1.2 or 1.4  $\text{M}_\odot$. Furthermore, a positive slope at   1.8 $\text{M}_\odot$ would suggest the presence of exotic matter within the NS.  

Within our sets, we could also conclude that a polytropic index below 1.75 or the trace anomaly related quantity $d_c<0.2$ does not necessarily indicate the presence of quark matter, as suggested in \cite{Annala_2020,Annala_2023}: some of our hybrid EOS predict $d_c>0.2$  in the quark phase, and the hadronic EOS may have $d_c<0.2$ at large densities. We have also analyzed the maximum NS compactness determined from our datasets. Values below 0.3 were obtained, consistent with the findings of \cite{Rezzolla:2025pft}.
   
    All sets can describe both theoretical constraints and observational data.
    However, NS data carry large uncertainties, especially regarding the radius values.
    This makes it difficult to make strong statements about the matter phase of their inner cores.
    Third-generation telescopes are expected to measure radii within $< 100$ meters \cite{cosmicexplorer},
    which would impose strong constraints on the EOS.
    This could provide more information about the composition of matter and its properties
    under the extreme conditions that are uniquely reproduced by NSs.

\section*{Acknowledgments}
MA expresses sincere gratitude to the FCT for their generous support
through Ph.D. grant number 2022.11685.BD
(DOI: \hyperlink{https://doi.org/10.54499/2022.11685.BD}{https://doi.org/10.54499/2022.11685.BD}). 
This research received partial funding from national sources through the FCT (Fundação para a Ciência e a Tecnologia, I.P, Portugal) for project UID/04564/2025 identified by DOI 10.54499/UIDB/04564/2025.  This work was supported by computational resources from the Deucalion HPC system in Portugal under Advanced Computing Project 2025.00067.CPCA A3, part of the National Advanced Computing Network (RNCA - Rede Nacional de Computação Avançada), funded by the Portuguese Foundation for Science and Technology (FCT - Fundação para a Ciência e a Tecnologia, IP).

\section*{Data Availability}
The final posterior of the model parameters,
the equation of states and the TOV results of the
NJL, MFTQCD, RMF, NJL-GW and r-NJL sets can be obtained from the link
\hyperlink{https://doi.org/10.5281/zenodo.17534438}{https://doi.org/10.5281/zenodo.17534438}.

\bibliography{refs}

\appendix

\section{Bayesian Approach Details} \label{sec:Bayes_Details}
    In this section, we present all the likelihoods applied in this work.
    We used 10 different constraints, which we divided them into 3 groups:
    experimental/observational data,
    guaranteeing a hybrid EOS, and
    corrections at M$_\text{max}$.
    
    \subsection*{1. Experimental/observational data}
        \noindent
        \textbf{Nuclear matter properties ($\mathcal{L}_\text{NMP}$)}
           constrain the EOS to satisfy the Nuclear matter properties (NMP)
            presented in Table \ref{tab:nmp_constraints} with the log-likelihood
            \begin{equation}
                \log(\mathcal{L}_\text{NMP}) =
                    -\frac{1}{2} \sum_j \left[
                        \left( \frac{d_j -m_j(\theta)}{\sigma_j} \right)^2
                        +\log(2 \pi \sigma_j^2)
                    \right].
            \end{equation}
      The first restriction in Table \ref{tab:nmp_constraints}
      concerns the saturation density ($\rho_0$).
      This is defined as the density of the symmetric nuclear matter
      at which the binding energy reaches its minimum, i.e.
      \begin{equation}
        \left. \frac{\partial (EA)}{\partial \rho} \right|_{\rho=\rho_0} = 0,
      \end{equation}
      where $EA = \epsilon/\rho -m_n$ with $m_n = 939 \text{ MeV}$.
      The second restriction comes from the known values of $EA(\rho = \rho_0)$
      and the incompressibility
      \begin{equation}
        K_0 = 9 \rho_0^2 
        \left. \frac{\partial^2 (EA)}{\partial \rho^2} \right|_{\rho=\rho_0},
      \end{equation}
      also for symmetric nuclear matter.
      The last one is applied to the symmetry energy at saturation,
      \begin{equation}
        J_\text{sym,0} = S(\rho_0) = \frac{1}{2} 
        \left. \frac{\partial^2 (EA)}{\partial \delta^2} \right|_{\delta=0},
      \end{equation}
      where $\delta = (\rho_n -\rho_p)/\rho$ is the isospin asymmetry
      and 
      {the second derivative with respect to $\delta$ is taken at $\delta=0$, i.e. for symmetric matter }.

      It is also possible to calculate the skewness $Q_0$ ($n = 3$)
      and kurtosis $Z_0$ ($n = 4$) coefficients using
      \begin{equation}
      \label{eq:X_nmp}
        X_0^{(n)} = 3^n \rho_0^n
        \left(
          \frac{\partial^n EA}{\partial \rho^n}
        \right)_{\delta = 0},
      \end{equation}
      and the symmetry energy slope $L_{\text{sym},0}$ ($n = 1$),
       curvature $K_{\text{sym},0}$ ($n = 2$),
       skewness $Q_{\text{sym},0}$ ($n = 3$), and
       kurtosis $Z_{\text{sym},0}$ ($n = 4$), given by
      \begin{equation}
      \label{eq:X_nmp_sym}
        X_{\text{sym},0}^{(n)} = 3^n \rho_0^n
        \left(
          \frac{\partial^n S(\rho)}{\partial \rho^n}
        \right)_{\rho_0}.
      \end{equation} \\

    \noindent
    \textbf{Pure Nuclear Matter ($\mathcal{L}_\text{PNM}$):}
      constrains the EOS to satisfy the pure nuclear matter (PNM) {energy per neutron} from $\chi$EFT shown in Table
      \ref{tab:nmp_constraints} with the log-likelihood
      \begin{equation}
        \log(\mathcal{L}_\text{PNM}) =
          \log
          \left[
            \prod_j \frac{1}{2 \sigma_j}
            \frac{1}{\exp{\frac{|d_j -m_j(\theta)| -\sigma_j}{0.015}}+1}
          \right].
      \end{equation} 

        \noindent
        \textbf{X-ray NICER Data ($\mathcal{L}_\text{NICER}$):}
           The EOS must be able to describe observational data.
           Here we use the NICER data from
           J0030+0451 \cite{Vinciguerra_2024,Miller_2019} (with the ST+PDT hotspot model),
           J0740+6630 \cite{Salmi_2024,Miller:2021qha} and
           J0437+4715 \cite{Choudhury_2024,Reardon_2024}.
           The likelihood is given by
           \begin{align}
               P(d_{X-ray} | EOS) &=
                \int_{M_\text{min}}^{M_\text{max}} dm P(m|EOS) \nonumber \\
                &\quad \times P(d_{X-ray} | m, R(m,EOS)) \nonumber \\
                &= \mathcal{L}_\text{NICER},
           \end{align}
           where
           \begin{equation}
               P(m|EOS) = 
               \begin{cases}
                   \frac{1}{M_\text{max} -M_\text{min}}, \text{ if } M_\text{min} \leq m \leq M_\text{max}, \\
                   0, \text{otherwise},
               \end{cases}
           \end{equation}
           with $M_\text{min} = 1 \text{M}_\odot$ and $M_\text{max}$ the maximum mass obtained from the EOS. \\
           
        \noindent
        \textbf{Gravitational Wave Data ($\mathcal{L}_\text{GW}$):}
        We use the GW170817 data \cite{Abbott_2018} from the LIGO-Virgo Collaboration.
        The likelihood is given by
        \begin{equation}
            \mathcal{L}_\text{GW} =
            \prod_i P(\Lambda_{1,i}, \Lambda_{2, i}, q_i |
                \mathcal{M}_c, \boldsymbol{d}_{GW,i} (\boldsymbol{d}_{EM,i})),
                \label{eq:gw}
        \end{equation}
        where $\Lambda_{j,i}$ is the tidal deformability of the $j$ binary component,
        $q_i$  the mass ratio,
        $\mathcal{M}_c$ the chirp mass, and $\boldsymbol{d}_{GW,i}$  the observational data.
        The chirp mass is fixed at $\mathcal{M}_c = 1.186 \text{M}_\odot$.
           
    \subsection*{2. Ensuring a hybrid EOS}
        \noindent
        \textbf{Minimum distance between hadron and quark $P \times \mu_B$ ($\mathcal{L}_\text{dist}$):}
		      To generate a hybrid EOS through Maxwell construction,
		      the quark and hadron equations should have an intersection point in the
		      $P \times \mu_B$ plot.
		      However, depending on the parameter values,
		      there may be cases in which the $P_Q(\mu)$ and $P_H(\mu)$ curves do not intersect,
		      and a phase transition is impossible.
		      To ensure that we have this intersection point,
		      we apply the following likelihood:
		      \begin{equation}
		        \log(\mathcal{L}_\text{dist}) = -\frac{x^2}{2 \delta^2},
		      \end{equation}
		      where $x$ is the minimum distance between the $P_Q(\mu)$ and $P_H(\mu)$ curves
		      and $\delta = 0.01$.
		      $\mathcal{L}_\text{dist}$ is a narrow Gaussian centered at $x = 0$.
		      The smaller the distance between the quark and hadron curves,
		      the closer $\log(\mathcal{L}_\text{dist})$ is to zero.\\

\begin{figure}[ht]
			     \centering
			     \includegraphics[width=.9\linewidth]{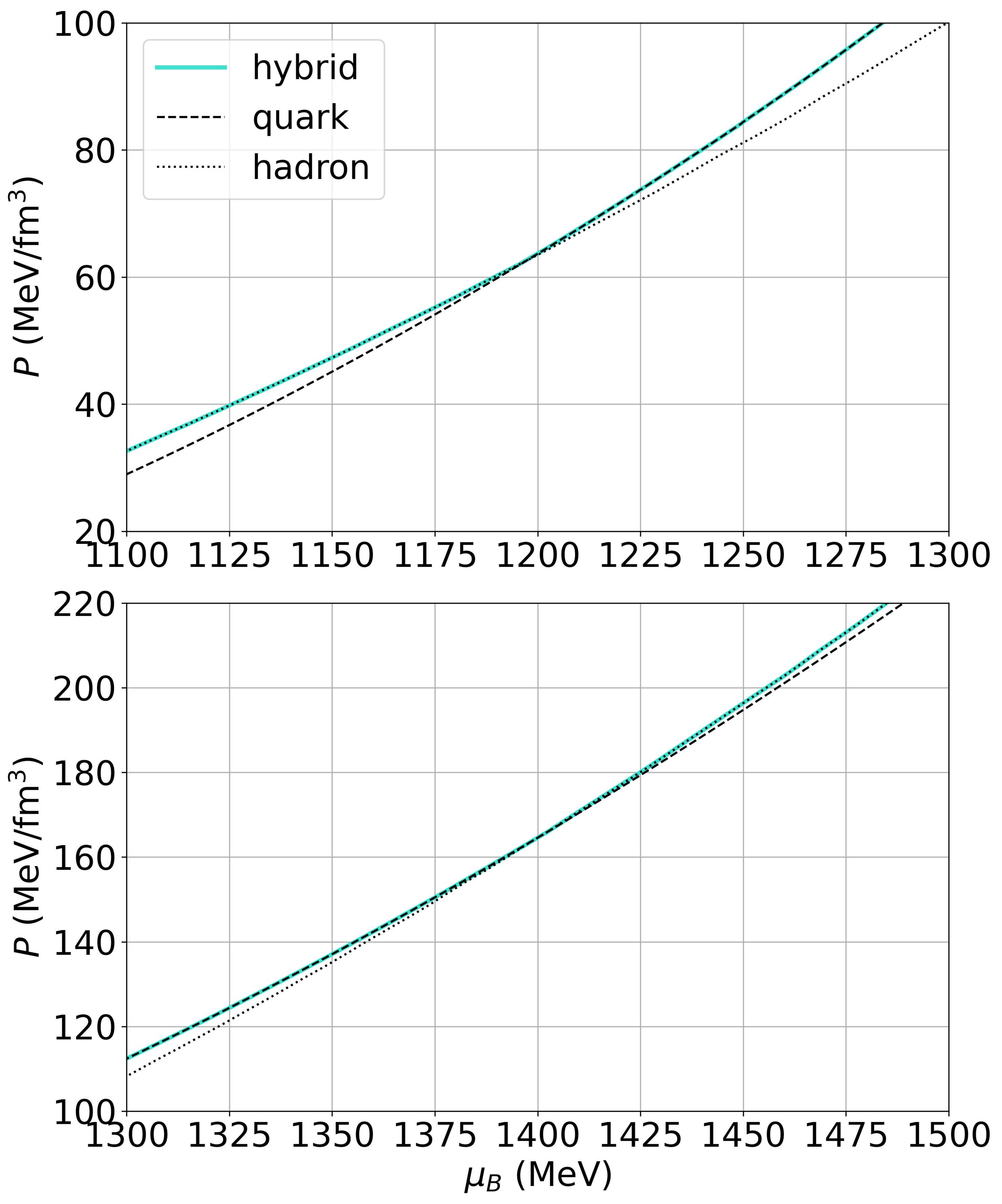}
			     \caption{Example of a physical (top) and non-physical (bottom) phase transition.
                 Cyan solid, black dashed and black dotted lines represent
                the hybrid, quark and hadron equations.}
			     \label{fig:paper2_HtoQ}
			\end{figure}

        \noindent
        \textbf{Phase transition from hadron to quark ($\mathcal{L}_\text{HtoQ}$):}
			In Maxwell's construction, for each value of $\mu$,
			the pressure value will be the largest value between $P_H(\mu)$ and $P_Q(\mu)$.
			If $P_H(\mu) > P_Q(\mu)$ ($P_Q(\mu) > P_H(\mu)$),
			this indicates that the matter is in the  hadron (quark) phase  at $\mu$.
			A hybrid equation can thus be represented by the top plot of
			Fig. \ref{fig:paper2_HtoQ}.
			Therefore, we will have hadron matter at $\mu < \mu_\text{trans}$,
			and quarks at $\mu > \mu_\text{trans}$.
			However, depending on the values of the parameters,
			we can have the case represented by the bottom plot of Fig. \ref{fig:paper2_HtoQ},
			resulting in a non-physical hybrid equation.
			To avoid this situation, we used the following likelihood:
			\begin{equation}
			  \mathcal{L}_\text{HtoQ} = \frac{1}{1+\exp(ax+b)},
			  \label{eq:step}
			\end{equation}
			where $a = -6$, $b = 1.5$, $x = P_H (\mu_0) -P_Q (\mu_0)$
			with $\mu_0$ being a small value (in this case, $\mu_0 = \mu_Q (\rho_B = 0.235)$).
			This equation behaves as a smooth step function.
			Adjusting the $a$ and $b$ parameters,
			we can set the position of the step and its smoothness.
			Here, we choose $a$ and $b$ values to result in
			$\mathcal{L}_\text{HtoQ} \approx 1$ when
			$P_H (\mu_0) > P_Q (\mu_0)$ (hadron phase at $\mu_0$)
			and
			$\mathcal{L}_\text{HtoQ} \approx 0$ in the opposite case.\\
			
        \noindent
        \textbf{Range of phase transition ($\mathcal{L}_\text{phT}$):}
            We constrained the density of the phase transition value using a super-gaussian
            centered at $\rho = 0.275\, \text{fm}^{-3}$ with a standard deviation $\sigma = 0.08$ and $p = 5$.
            The likelihood is written as
            \begin{equation}
                \log(\mathcal{L}_\text{phT}) = -\left[
                    \frac{(\rho_\text{trans} -0.275)^2}{2(0.08)^2}
                \right]^5.
            \end{equation}
            These chosen values imply that $\rho_0 \lesssim \rho_\text{trans} \lesssim 0.40$.
        
    \subsection*{3. Corrections at M$_\text{max}$}
\begin{figure}[ht]
			     \centering
			     \includegraphics[width=\linewidth]{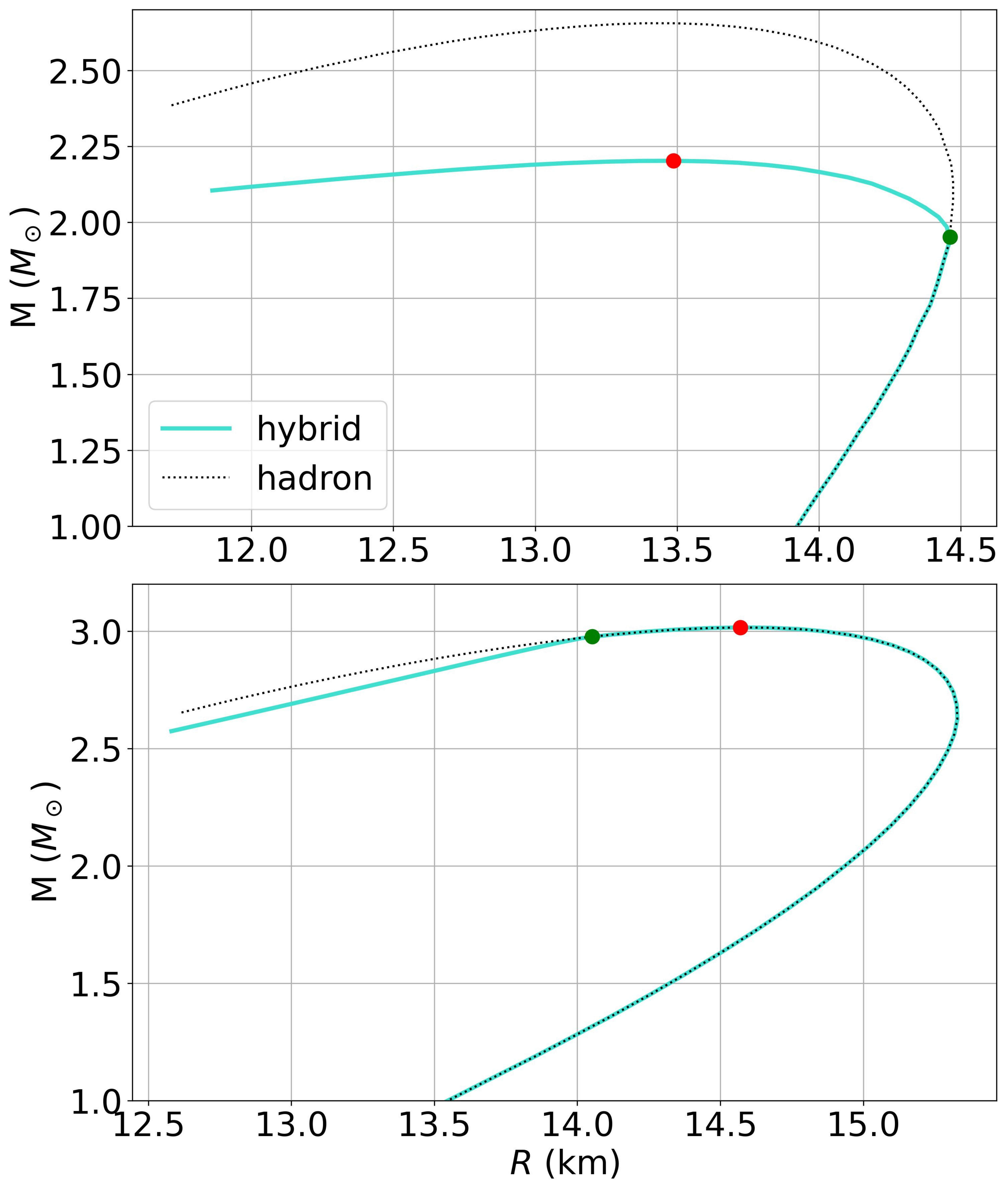}
			     \caption{Example of a physical (top) and non-physical (bottom) phase transition.
                 The red dot shows the maximum mass,
                 and the green dot shows the phase transition.}
			     \label{fig:paper2_Pmax}
			\end{figure}

        \noindent
        \textbf{Quarks inside M$_\text{max}$ ($\mathcal{L}_\text{Qmax}$):}
	        In this study, we are interested in exploring cases in which
	        quark matter can be found in stable NSs.
	        To obtain these cases, the constraint $P_\text{max} > P_\text{trans}$
	        needs to be imposed.
	        Fig.  \ref{fig:paper2_Pmax} shows the two possible
	        hybrid mass-radius diagrams.
	        The top plot shows the case in which we are interested,
	        and the bottom plot shows the case we are avoiding.
	        In the first case, NSs with $M > 2 \text{M}_\odot$ have quark matter in their core.
	        In the second case, the maximum mass does not have enough pressure to deconfine matter.
	        Hybrid stars are only possible in the unstable branch.
	        To ensure that the EOS results in a maximum-mass NS
	        with a significant amount of quark matter inside it,
	        we used the smooth step function from Eq. (\ref{eq:step}) with
	        $a = -0.2$, $b = 20$, and $x = P_\text{max} -P_\text{trans}$.
	        These values were chosen to position the step at approximately 100.
            \\ 

        \noindent
        \textbf{Causality at M$_\text{max}$ ($\mathcal{L}_{c_s^2\text{max}}$):}
            To guarantee causality inside NSs,
            we impose that $c_s^2 < 1$ at the central density of $M_\text{max}$.
            To do this, we define the likelihood as a smooth step function, Eq. (\ref{eq:step}), with the values $a = 100$, $b = -92$ and $x = c_s^2 (\rho_\text{max})$.\\

        \noindent
        \textbf{Perturbative QCD ($\mathcal{L}_\text{pQCD}$):}
        In \cite{Gorda_2023},
        the authors developed a code \cite{komoltsev_2023_7781233}
        in which a Monte Carlo integration is performed for a given point of
        pressure, energy density, and baryonic density ($P, \epsilon, \rho_B$)
        to verify whether this point satisfies the pQCD constraint
        for an energy scale $X = [1/2, 2]$.
        This code is based on the pQCD constraints discussed in \cite{Komoltsev:2021jzg}.
        The authors developed a method that constrains the
        $P \times \epsilon \times \rho_B$ space
        based solely on thermodynamic relations and ab initio calculations
        ($\chi$EFT \cite{Hebeler_2013} and pQCD \cite{PhysRevLett.127.162003}).

    \subsection*{Full Likelihood}
    For the NJL, r-NJL and MFTQCD sets, the total likelihood can be written as
    \begin{align}
      \log(\mathcal{L}) &=
         \log(\mathcal{L}_\text{NMP})
        +\log(\mathcal{L}_\text{PNM})
        +\log(\mathcal{L}_\text{NICER})
        \nonumber \\
        &\quad
        +\log(\mathcal{L}_\text{dist})
        +\log(\mathcal{L}_\text{HtoQ})
        +\log(\mathcal{L}_\text{phT})
        \nonumber \\
        &\quad
        +\log(\mathcal{L}_\text{Qmax})
        +\log(\mathcal{L}_{c^2_s\text{max}})
        \nonumber \\
        &\quad
        +\log[\mathcal{L}_\text{pQCD}(7 \rho_0)],
    \end{align}
    where we applied the pQCD constraint $\mathcal{L}_\text{pQCD}$ at $\rho_B = 7 \rho_0$.
    For the RMF set, the total likelihood is
    \begin{align}
      \log(\mathcal{L}) &=
         \log(\mathcal{L}_\text{NMP}) 
        +\log(\mathcal{L}_\text{PNM})
        +\log(\mathcal{L}_\text{NICER})
        \nonumber \\
        &\quad
        +\log(\mathcal{L}_{c^2_s\text{max}})
        +\log[\mathcal{L}_\text{pQCD}(7 \rho_0)].
    \end{align}
    For the NJL-GW set, 
    we applied the observation data GW170817 ($\mathcal{L}_\text{GW}$):
    \begin{align}
      \log(\mathcal{L}) &=
         \log(\mathcal{L}_\text{NMP})
        +\log(\mathcal{L}_\text{PNM})
        \nonumber \\
        &\quad
        +\log(\mathcal{L}_\text{NICER})
        +\log(\mathcal{L}_\text{GW})
        \nonumber \\
        &\quad
        +\log(\mathcal{L}_\text{dist})
        +\log(\mathcal{L}_\text{HtoQ})
        +\log(\mathcal{L}_\text{phT})
        \nonumber \\
        &\quad
        +\log(\mathcal{L}_\text{Qmax})
        +\log[\mathcal{L}_\text{pQCD}(7 \rho_0)].
    \end{align}
    The value of $7 \rho_0$ for the pQCD constraint was chosen due to the fact that
    the central densities at the maximum NS mass
    for the hybrid EOS sets can reach only {$\sim 7 \rho_0$}
    (see Table \ref{tab:NS_prop} {and \ref{tab:NS_prop_RMF}}).

  \section{Hadron parameters}
  
  \begin{figure}[ht]
    \centering
    \includegraphics[width=.9\linewidth]{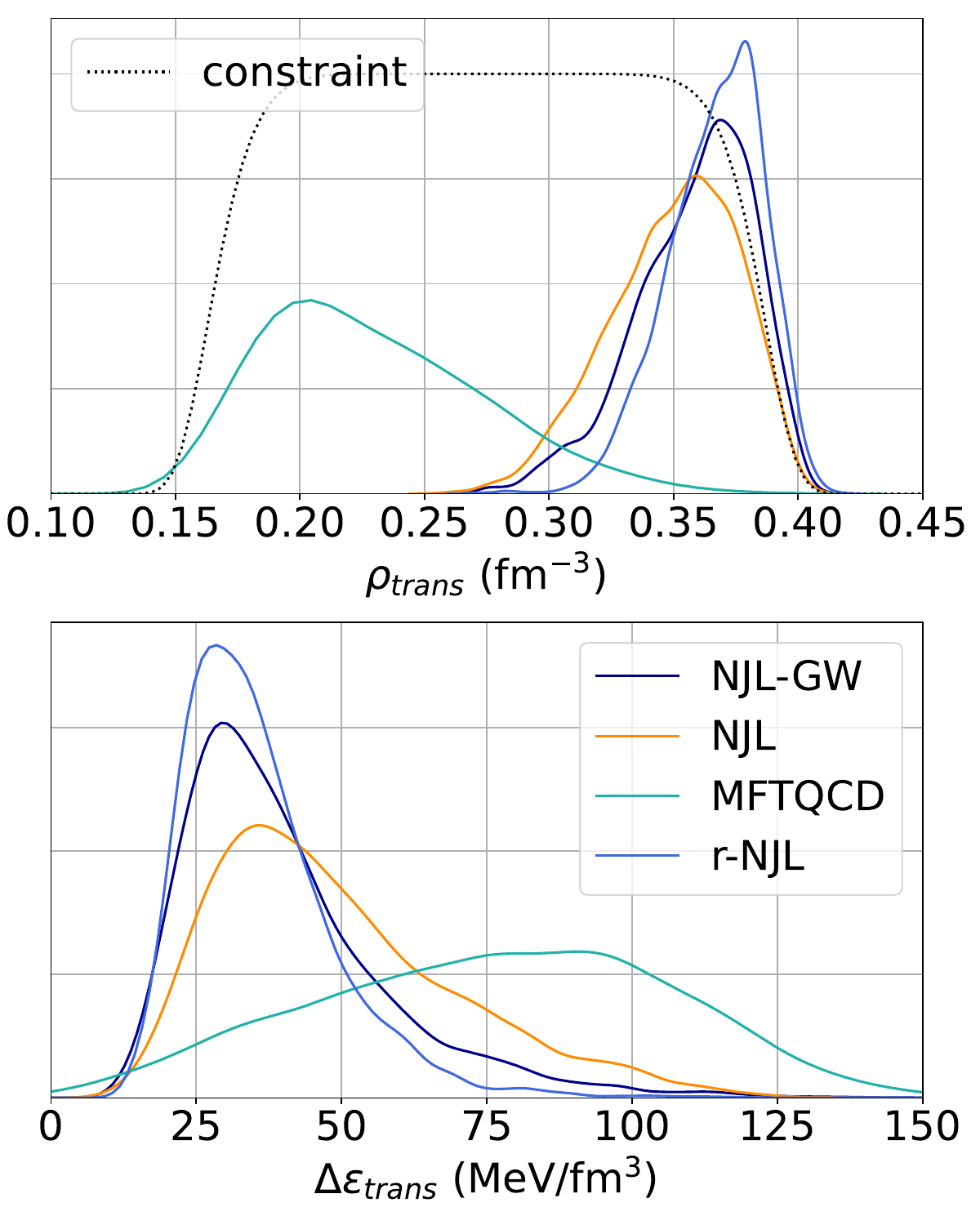}
    \caption{Corner plot of the hadron parameters for sets NJL, MFTQCD and RMF.
      Same color code from Fig. \ref{fig:paper2_PxE}.}
    \label{fig:corner_H}
  \end{figure}
  
  Fig. \ref{fig:corner_H} shows the corner plot of the hadron parameters.
  It can be seen that the MFTQCD model does not significantly change
  the values of the hadron parameters compared to the RMF set,
  except for the $\xi$ parameter, to which the model is not sensitive. This is because this term plays a role at high densities when the phase transition to quark matter has already occurred.
  Within the  NJL model, a different result was obtained: the values of $g_\sigma$, $g_\omega$ are much larger, and there is no superposition with the values obtained for RMF and MFTQCD;  $g_\rho$  shows a wider distribution and may take larger values than those of RMF and MFTQCD;   BB, CC and $\Lambda$ peak at smaller values.
  $\xi$ is the only parameter for which its posterior remains approximately the same.
  The differences in the parameter distributions explain why the NJL set has larger radii.  It is a consequence of imposing that two solar mass stars are described and that the hybrid star has a non-negligible quark core. To attain these conditions, the hadron EOS must be quite stiff to allow an early transition to quark matter. 
  However, one consequence is that the NJL set does not satisfy the GW170817 constraints, as can be seen in Fig. \ref{fig:paper2_tidal}. This is not the case with the MFTQCD and RMF sets, which satisfy the GW170817 constraints, although they have not been included in the Bayesian inference.
    \\

 \section{Quark-Hadron Parameters Relation}
  
\begin{figure}[hb]
    \centering
    \includegraphics[width=.9\linewidth]{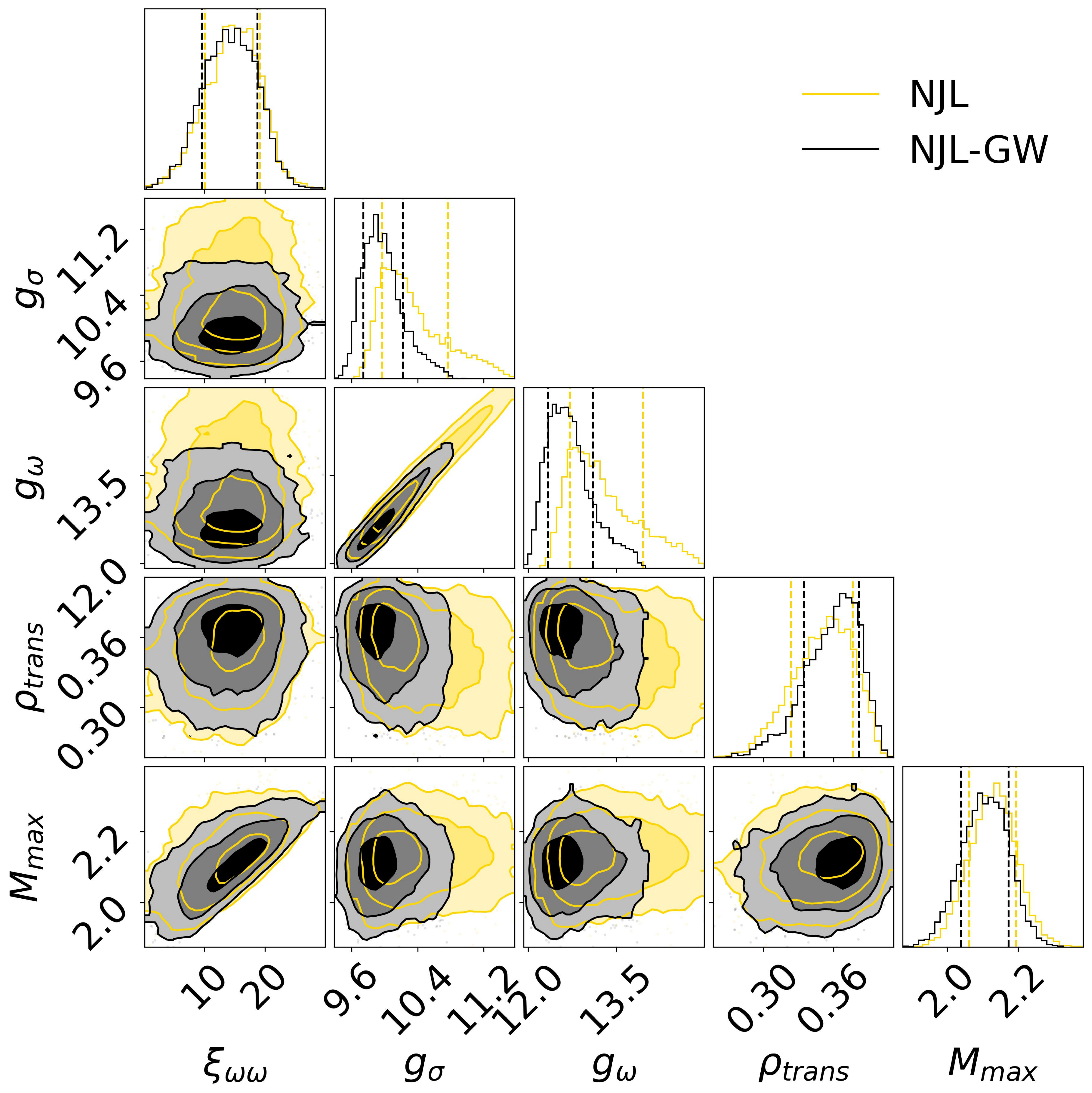}
    \caption{Corner plot of $\xi_{\omega\omega}$, $g_\sigma$, $g_\omega$,
      $\rho_\text{trans}$ and $M_\text{max}$ for the NJL sets with different priors.}
    \label{fig:paper2_corner_prop_prior}
  \end{figure}

  Even when applying the GW170817 constraint or restricting the hadron parameters prior,
  it is notable that NJL-GW and r-NJL sets radius are still larger than those of the MFTQCD and RMF sets.
  At this point, we might ask,
  ``Why do Bayesian analyzes prefer EOS with larger radii for the NJL sets?''
  Larger values of $g_\omega$,
  which increase the radius, decrease the value of $\xi_{\omega\omega}$
  (see Fig. \ref{fig:paper2_corner_prop_prior})
  and, therefore, that of the transition density $\rho_\text{trans}$.
  This allows for larger values of the $\xi_{\omega\omega}$ coupling constant,
  which increases the maximum mass.
  In other words, increasing $g_\omega$ offsets the increase in $\xi_{\omega\omega}$,
  resulting in values of $\rho_\text{trans}$ within the Bayesian constraint.
  Additionally, current observational data set stronger constraints on the mass than on the radius,
  implying that Bayesian inference will prioritize larger $M_{\text{max}}$ than $R_{1.4} \sim 13$ km.
  
\end{document}